\documentclass[12pt,onecolumn]{article}

\usepackage[left=38.5pt,%
right=43pt,%
top=53pt,%
bottom=57pt]{geometry}

\usepackage{amsmath,amssymb,amsfonts,epsfig,color}
\usepackage{threeparttable}

\usepackage[
backend=biber,
style=nature, 
sorting=none,        
giveninits=true,     
isbn=false,          
doi=true             
]{biblatex}
\AtBeginBibliography{\small} 

\usepackage{authblk}

\usepackage[mathlines]{lineno}

\usepackage{setspace}

\usepackage{booktabs} 
\usepackage{tablefootnote} 

\usepackage{amsthm}
\usepackage{epstopdf}
\usepackage{graphicx}
\usepackage[dvipsnames]{xcolor}
\usepackage{helvet}
\usepackage{layouts}
\usepackage{multirow}

\graphicspath{{./figure/}}

\def\IR{\mathbb{R}}

\renewcommand{\t}{^{\mbox{\tiny\sf T}}}
\newcommand{\hs}{\hspace{4mm}}
\newcommand{\red}{\color{black}}
\newenvironment{mat}{\left[\!\!\!\begin{array}}{\end{array}\!\!\!\right]}

\usepackage{multirow}%
\usepackage[makeroom]{cancel}
\usepackage{arydshln}

\title{\Large\bf Dynamics modeling and analysis of batoid-type locomotion powered by tensegrity wing structure}

\author[1]{\small Jun Chen \thanks{Corresponding author: jchen@lsnl.cn}}
\author[2]{\small Tetsuya Iwasaki}
\author[3]{\small Yuhong Liu \thanks{Corresponding author: yuhong\_liu@tju.edu.cn}}
\affil[1]{\small Department of Ocean Observation and Detection, Laoshan Laboratory, Qingdao 266237, China}
\affil[2]{\small Department of Mechanical and Aerospace Engineering, University of California, Los Angeles, CA 90095, USA}
\affil[3]{\small Key Laboratory of Mechanism Theory and Equipment Design of Ministry of Education, School of Mechanical Engineering, Tianjin University, Tianjin 300354, China}

\date{}

\begin{document}

\maketitle

\abstract{Control signals and kinematics in batoid swimming are difficult to measure experimentally, making body-fluid interaction models essential for studying their underlying locomotion principles. To address this challenge, we developed a body-fluid interaction model of batoid-type swimming that is appropriate for both neural control study and engineering design. The body trunk is modeled as a rigid body with six degrees of freedom. {\red The flexible pectoral fins attached to the trunk are modeled by a tensegrity structure consisting of rigid struts and elastic cables that resembles a biological musculoskeletal system. The fin is actuated by changing the tension of elastic cables distributed across the fin surface, enabling controllable and realistic deformation.
Utilizing an analytical fluid force model, the body-fluid interaction model is exercised through simulation examples that respectively investigate the speed difference between tension actuation and fin kinematic waves, the effects of fin stiffness and resonance exploitation on swimming performance, and the different fin kinematics resulting from different body inertial motions.}
}

\vspace{6pt}

{\noindent\bf Keywords:}  Modeling and dynamic analysis, Batoid swimming, Tensegrity structure, Bio-inspired underwater vehicle

\section{Introduction}\label{sec1}

{\red Batoid fishes, such as the manta ray and butterfly rays, propel themselves by two pectoral fins. These rays are able to migrate over long distances, swim fast at sustained speed 2.8~m/s, and maneuver within confined spaces or close to the seafloor. Different fin kinematics are observed among these rays: the fin kinematics of manta rays are more like the flapping motion, while the butterfly fishes show traveling waves on fin surface. Details of fin kinematics and actuation signals of the fin are difficult to measure due to their three-dimensional motion and the living environment for large-sized fishes such as manta rays. To study the generation of fin kinematics subjected to both muscle activation and fluid force and to study neural control principles, developing the model of body and pectoral fin dynamics is necessary.}

{\red Studies of anguilliform or carangiform fish swimming have shown the important role of body dynamics in swimming control.} For example, body kinematics, motor neuron activation signals, and tension have different traveling speeds along the body depending on the body stiffness, and the wave speeds of the three vary among species \cite{jchen:waves,williams:89shortcommunication}. Fishes actively tune muscle stiffness to match body resonance frequency to the hydrodynamic resonance to save energy \cite{kohannim:14,schwalbe:18,mchenry:95,long:96,long:98,tytell:16,triantafyllou:93}. The optimal value of {\red body} stiffness for maximum acceleration is different from that for maximum steady swimming speed \cite{tytell:10pnas}. Muscle dynamics affect the swimming gait \cite{hamlet:15}. The principles of gait adaptation via sensory feedback are studied by integrated leech \cite{iwasaki:pnas} and lamprey \cite{hamlet:18} models from central pattern generator to body-fluid interaction. {\red To perform similar analyses for batoid swimming, a model incorporating both the body and pectoral fin dynamics is required. However, studies on the dynamic modeling of flexible pectoral fins and free swimming of batoid fishes are very limited. Reference \cite{blair:thesis,liu:16} modeled the pectoral fin by small and thin plates interconnected by springs and dampers to calculate the optimal fin kinematics for various cost functions.}
Most studies {\red on batoid swimming} have focused on the hydrodynamics of pectoral fin propulsion \cite{fish:16,Dong:22_biomimetics,moored:11_CFD,Liu2015ThrustPM,xibei_JFM:2022}. The virtual skeleton-based 3D surface reconstruction in combination with video data is used to approximate pectoral fin kinematics. In computational fluid dynamic simulation, the body trunk and morphing fins are described by the moving boundary surfaces. {\red Such a simulation setting ignored the effects of fin inertia, stiffness, and damping on gait selection and swimming performance.}

{\red To best emulate biological pectoral fin, reference \cite{chen:21,moored:11} designed a tensegrity pectoral fin where individual fin rays are modeled by planar tensegrity beams, and elastic cables connect the adjacent tensegrity beams to emulate the tissue elasticity between fin rays. Tensegrity structure is composed of rigid struts and elastic cables, resembling biological musculoskeletal system. Each tensegrity beam is actuated by multiple cables along its length to simulate biological muscle functions \cite{bliss:13}. Such design of tensegrity fin has two advantages. Firstly, the struts and cables endure only axial forces, which gives it a high stiffness-to-mass ratio suitable for large-sized vehicles (such as the manta rays). Secondly, the three-dimensional deformation of the fin is achieved by changing the cable rest length within the planar tensegrity beam which makes the actuation mechanism relatively simple for engineering implementation \cite{bliss:13,moored:11}. In this paper, we develop the dynamic model of batoid swimming powered by such tensegrity fins. Based on this model, we investigate how fin kinematics are generated by tension actuation waves subjected to the fluid forces and how fin stiffness and natural frequencies affect swimming performance.}

The paper is organized as follows. In section~\ref{sec:modeling}, we develop the body-fluid interaction model of batoid swimming where the body trunk is modeled by a rigid body with six degrees of freedom and two tensegrity pectoral fins are attached to the body trunk. In section~\ref{sec:results}, an analytical fluid force model is used to approximate the fluid force (which model has been successfully used in the optimal gait study of batoid swimming in \cite{liu:16,blair:11}) {\red and free swimming is simulated. The straight-line swimming is first simulated to see what actuation signals are like to generate the chord- and span-wise traveling waves on the fin (section \ref{sec:wing_kinematics}). Then, the effects of fin stiffness and resonance exploitation on swimming performance are investigated (section \ref{sec:simulation_resonance}). In section \ref{sec:simulation_outOfPhase}, the straight-line swimming with out-of-phase flapping of two fins is simulated to compare the fin kinematics with those of in-phase flapping straight-line swimming. In section \ref{sec:simulation_attitude}, the somersault, rolling and turning motions are simulated by changing fin kinematics.} We summarize our work in the conclusion section \ref{sec:conclusion}.

\section{Dynamic model of tensegrity batoid body subjected to external forces}\label{sec:modeling}

\subsection{Description of tensegrity batoid body}
\begin{figure}[!t]
	\centering
	\begin{minipage}{0.5\textwidth}
		\includegraphics[width=\textwidth,trim=0mm 10mm 2mm 10mm,clip=true]{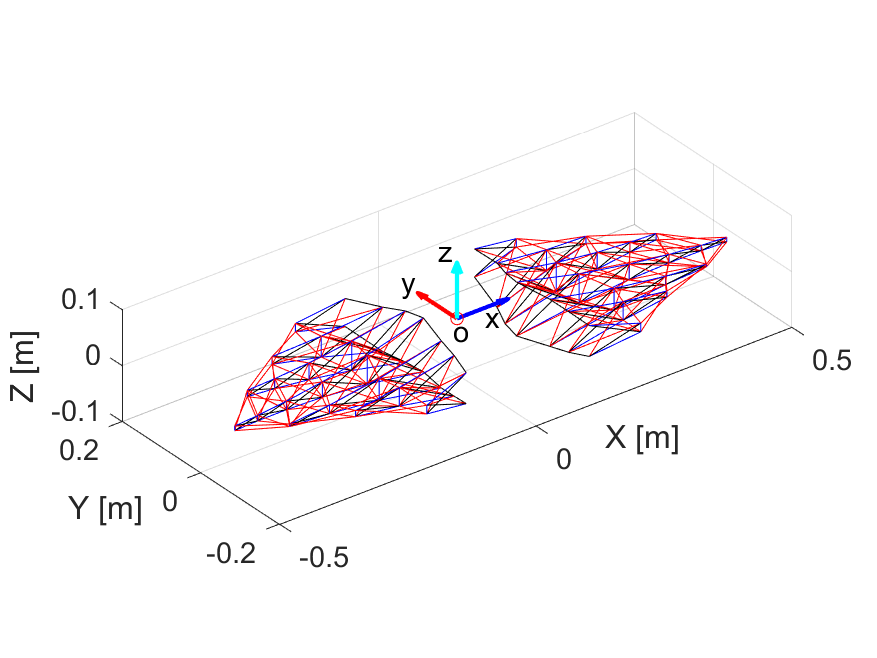}
	\end{minipage}
	\begin{minipage}{0.23\textwidth}
		\includegraphics[width=\textwidth,trim=0mm 0mm 0mm 0mm,clip=true]{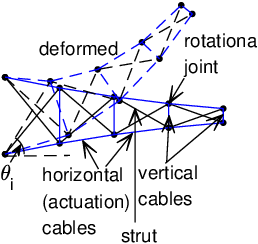}
	\end{minipage}
	\begin{minipage}{\textwidth}\hspace{5cm} (a) \hspace{3.8cm} (b)
	\end{minipage}\\
	\begin{minipage}{0.3\textwidth}
		\includegraphics[width=\textwidth,trim=0mm 0mm 5mm 0mm,clip=true]{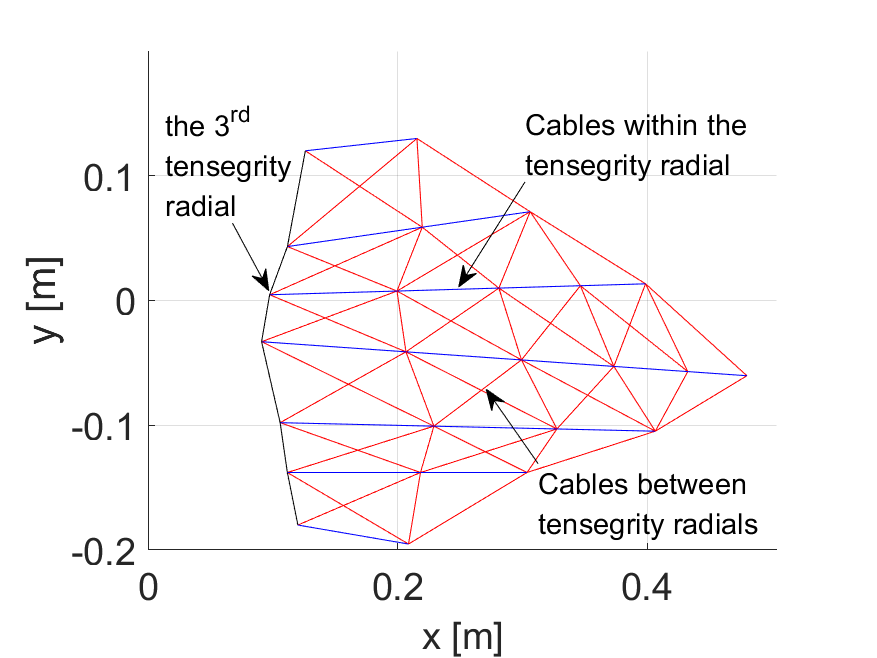}
	\end{minipage}
	\begin{minipage}{0.3\textwidth}
		\includegraphics[width=\textwidth,trim=35mm 40mm 35mm 30mm,clip=true]{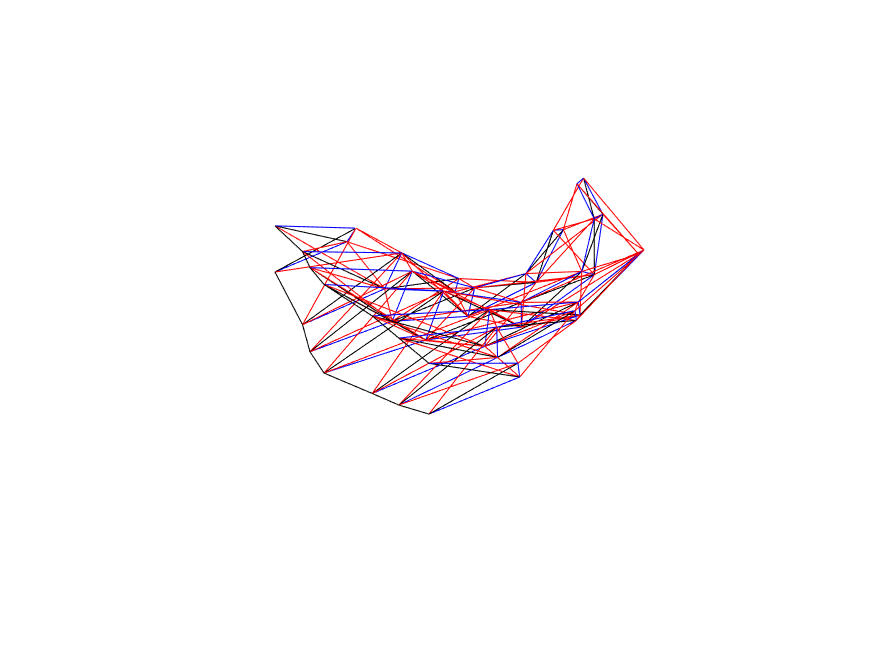}
	\end{minipage}
	\begin{minipage}{0.3\textwidth}
		\includegraphics[width=\textwidth,trim=45mm 35mm 40mm 35mm,clip=true]{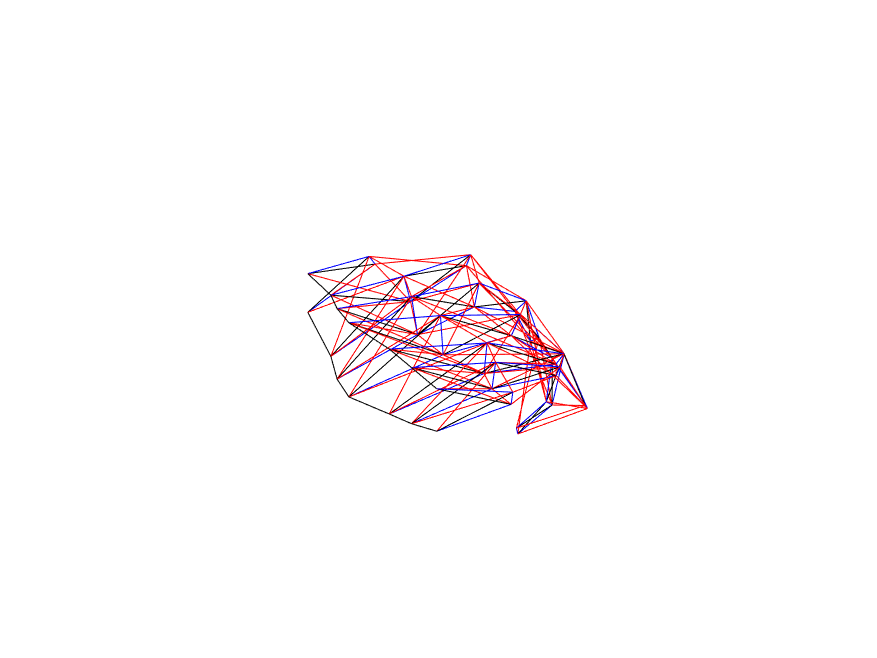}
	\end{minipage}
	\begin{minipage}{\textwidth}\hspace{2.3cm} (c) \hspace{3.2cm} (d) \hspace{3.4cm} (e)
	\end{minipage}
	\caption{(a) Configuration of a batoid type vehicle composed of two tensegrity wings and a rigid body trunk represented by the body frame $o-xyz$. The black lines represent the struts, and the blue and red lines respectively represent the elastic cables within and between tensegrity radials. (b) A tensegrity radial. (c) Top view of a single tensegrity wing. (d,e) Deformed wings subjected to the fluid force.}\label{fig:config_whole_body}
\end{figure}

Fig.~\ref{fig:config_whole_body}a shows a design of ray-like underwater vehicle with wings constructed by planar tensegrity radials (Fig.~\ref{fig:config_whole_body}b) and the body trunk represented by the body frame $o-xyz$. There are 7 radials in each wing and the $i^{th}$ radial is formed by a pair of strut chains, each of which consists of 
$n_i$ struts connected by rotational joints (Fig.~\ref{fig:config_whole_body}b, as shown in in Fig.~\ref{fig:config_whole_body}c $n_i=1,2,4,5,3,2,1$ respectively for the seven radials). The roots of the radials are attached to the rigid-body trunk through rotational joints. Between any two adjacent rotational joints within each radial, there is an elastic cable which is implemented by a spring connected in series between two rigid cables (i.e., $3n_i$ cables in the $i^{th}$ radial, Fig.1b).  Thus each radial consists of $n_i$ units, where each unit is formed by a pair of X-shaped struts connected by three cables.
Different from \cite{chen:21} where the tensegrity radials are parallel to each other, the orientation angles of radials are interpolated from biological data provided in \cite{blair:thesis} and the edge of the wing is equally spaced by these radials (Fig.~\ref{fig:config_whole_body}c).
We assume that the forces exerted by the cross-bridge cables (red lines in Fig.~\ref{fig:config_whole_body}a,c) on each tensegrity radial are approximately balanced and the planar radials only bend in their individual planes. By this approximation, the shapes of tensegrity radials and then the wing are described by strut angles with respect to the $x$-$y$ plane of the body frame ($\theta_i$ in Fig.~\ref{fig:config_whole_body}b, $i=1,\cdots,n-1$ and $n$ is the total number of rigid bodies including the struts in two wings and a single body trunk). The geometry of the wing is taken from the cownose ray \cite{blair:thesis} and scaled to the desired size 1.1~m between two wing tips.

Studies on eel-type swimming suggest that the posterior body is passive and body elasticity is utilized to transmit the power from mid-body to the posterior region where most thrust is generated \cite{jchen:fluid,tytell:16}. Similarly, we only actuate the first four anterior radials and leave the last three posterior radials passive. Also, not all the units in the first four radials are actuated; the numbers of actuated units for the first four radials are respectively 1, 2, 3 and 4, and one
unit near the wing tip for each of the third and fourth radials is not actuated. The entire system has 78 ($=36$ (for each wing) $\times 2+6$ (for body)) DOF within 40 actuators. Each of the horizontal actuation cables (40 in total for two wings) is connected to a motor. The elastic cable is composed of a spring connected in series between two rigid cables, and the length change of the rigid cable, rolled in or released by motor, changes the stretched length and tension of the spring to cause the radial deform (Fig.~\ref{fig:config_whole_body}b).
The wing skeleton is covered by elastic silicone and the encapsulated wing is assumed to have the same density as the fluid and neutrally buoyant.

\subsection{Equation of motion of tensegrity batoid body}

The batoid type underwater vehicle shown in Fig.~\ref{fig:config_whole_body} has $n$ rigid bodies ($n-1$ struts and one body trunk) and $n+5$ DOF. The body frame $o-xyz$ is attached to the body trunk. The generalized coordinates $q=\begin{bmatrix}w^\intercal & \phi^\intercal & \theta^\intercal \end{bmatrix}^\intercal$ are given by the position of the mass center of the whole system in the inertial frame $w\in\IR^3$, the orientation of the body frame with respect to the inertial frame $O-XYZ$, $\phi\in\IR^3$, and the body shape $\theta\in\IR^{n-1}$ (i.e., the strut angles of tensegrity wings with respect to the $x-y$ plane of the body frame). The general equation of motion {\red for tensegrity batoid body shown in Fig.~\ref{fig:config_whole_body}a} is given by (\ref{eqn:eom_theta}) and (\ref{eqn:eom_w}). The derivation of (\ref{eqn:eom_theta}) and (\ref{eqn:eom_w}), and the evaluation for coefficient matrices $\mathbb{J}(\vartheta)$ and $G(\vartheta,\dot\vartheta)$ are given in Appendix~\ref{appendix:general_eom} and Appendix~\ref{appendix:body_dynamics}.

\begin{align}
	& \mathbb{J}(\vartheta)\ddot\vartheta + G(\vartheta,\dot\vartheta)\dot\vartheta = h_\vartheta,  \label{eqn:eom_theta} \\
	& m_o\ddot w = h_w. 
	\label{eqn:eom_w}
\end{align}
where $\vartheta=\begin{bmatrix}\phi^\intercal & \theta^\intercal\end{bmatrix}^\intercal$,
\begin{equation}\label{eqn:J_vartheta_o}
	\mathbb{J}(\vartheta) = \sum_{i=1}^n m_i\left(\frac{\partial p_i}{\partial\vartheta}\right)\left(\frac{\partial p_i}{\partial\vartheta}\right)^\intercal + P_i(\vartheta)^\intercal J_iP_i(\vartheta),
\end{equation}
\begin{equation}\label{eqn:G_vartheta}
	G(\vartheta,\dot\vartheta)=\left(\frac{\partial\mathbb{J}(\vartheta)\dot\vartheta}{\partial\vartheta}\right)^\intercal - \frac{1}{2}\left(\frac{\partial\mathbb{J}(\vartheta)\dot\vartheta}{\partial\vartheta}\right),
\end{equation}
$p_i(\vartheta)$ and $P_i(\vartheta)$ in $\mathbb{J}(\vartheta)$ are functions that relate the position of the mass center of the $i^{th}$ rigid body in the inertial frame ($r_i\in\IR^3$) and its angular velocity expressed in the body frame $\omega_i\in\IR^3$ to the general coordinates, i.e.,
\begin{equation}\label{eqn:ri}
	r_i = w+p_i(\vartheta), \hs \omega_i = P_i(\vartheta)\dot\vartheta,
\end{equation}
$J_i$ is the moment of inertia around the mass center, $h_\vartheta$ and $h_w$ are the generalized forces, including the elastic cable force within tensegrity wings and fluid force, and will be given in the next two sections.

\subsection{Fin actuation and generalized force from cables}
{\red Fins are actuated by changing the rest length of horizontal elastic cables (\ref{fig:config_whole_body}b).} The generalized forces $h_\vartheta$ in (\ref{eqn:eom_theta}) includes both elastic force of cables  $\varphi(\vartheta)$ and fluid force. $\varphi(\vartheta)$ is derived from virtual work principle:
\begin{equation}\label{eqn:cable_force}
	\tau(\theta)^\intercal\delta l(\theta) = \varphi(\vartheta)^\intercal\delta\vartheta \hs \Rightarrow \hs \varphi(\vartheta) = \frac{\partial l(\theta)}{\partial\vartheta}\tau (\theta) =\begin{bmatrix}\bf{0} \\ \displaystyle\frac{\partial l(\theta)}{\partial\theta}\end{bmatrix}\tau (\theta)
\end{equation}
where $\tau(\theta)$ and $l(\theta)$ are respectively the cable force and length vectors, and they are only functions of wing shape $\theta$, the $i^{th}$ element of $\tau$ is given by
\begin{equation}\label{eqn:tau}
	\tau_i=-k_i\max(\Delta l_i,0)+c_i\dot{l}_i(\theta)\mathrm{sign}(\max(\Delta l_i,0)), \hs \Delta l_i = l_i(\theta)-(l_{i_0}+u_i),
\end{equation}
where $k_i$ and $c_i$ are respectively the cable stiffness and damping, $l_i(\theta)$ ($i=1,\cdots,n_c$) is the $i^{th}$ element of $l(
\theta)$, $l_{i_0}$ is the cable rest length at the initial equilibrium, $u_i$ is the control input to change the cable rest length $l_{i_0}$. The $i^{th}$ cable length $l_i(\theta)$ is given by
\[
l_i(\theta)=\sqrt{s_i(\theta)^\intercal s_i(\theta)},
\]
where $s_i\in\IR^3$ is the vector of the $i^{th}$ cable and given by the $i^{th}$ block of vector
\begin{equation}\label{eqn:cable_vector}
	s(\theta)=(C_s\otimes I_3)g^b(\theta) - g^b_o,
\end{equation}
where $g^b(\theta)\in\IR^{3(n-1)}$ is the vector of coordinates of strut ends with respect to the origin of the body frame, $C_s$ is called the connecting matrix describing the cables that connect between strut ends \cite{cheong:14,chen_nondyn}; for example, the $i^{th}$ row of $C_s$ defines the $i^{th}$ cable which connects the numbered $j^{th}$ and $k^{th}$ strut ends $g_j^b(\theta)\in\IR^{3}$ and $g_k^b(\theta)\in\IR^{3}$ which are respectively the $i^{th}$ and $j^{th}$ block element of $g^b(\theta)$; therefore, the $j^{th}$ and $k^{th}$ elements of the $i^{th}$ row of $C_s$ are given by $C_{s_{i,j}}=1$, $C_{s_{i,k}}=-1$ with the rest elements of the $i^{th}$ row of $C_s$ being zeros; vector $g_o^b$ is associated with those cables whose one end is connected to the body wall and given as follows: the $i^{th}$ block of $g_o^b$ is a 3 by 1 zero vector if the $i^{th}$ cable does not connect to the body wall or the body wall coordinate with respect to the origin of the body frame otherwise, $\otimes$ means multiplying each element of $C_s$ by a 3 by 3 identity matrix $I_3$, $n-1$ is the total number of struts. 
The expression for $g^b(\theta)$ in (\ref{eqn:cable_vector}) and the evaluation of $\partial l(\theta)/\partial\theta$ in (\ref{eqn:cable_force}) are given in Appendix~\ref{appendix:generalized_force}.

\subsection{Generalized force from fluid}
%
To be compatible with multiple rigid body model, the continuously distributed fluid forces on body trunk and wing surfaces are discretized and respectively act on the mass center of body trunk and the strut ends in tensegrity radials (the places where the silicone wing cover is supported). Let $f$ be the vector formed by stacking the discretized fluid forces $f_i\in\IR^3$ ($i=1,\cdots,n$) in a column and $g$ the vector of coordinates where the fluid forces $f_i$ are applied, both coordinatized in the inertial frame. The generalized fluid force vector $\varphi_f(\vartheta)$ is derived from virtual work principle:
\begin{equation}\label{eqn:fluid_force}
	\delta g(q)^\intercal f(q) =\delta q^\intercal \varphi_f(q)  \hs \Rightarrow \hs \varphi_f(q) = \frac{\partial g(q)}{\partial q}  f(q)
\end{equation}
The $i^{th}$ element of $g$ is given by
\begin{equation}\label{eqn:gi}
	g_i(q)= w + \Omega(\phi)^\intercal (g_i^b(\theta)-w^b(\theta)),
\end{equation}
where $\Omega(\phi)$ is the orthogonal rotation matrix that transforms the vector from the inertial frame to the body frame, $g_i^b(\theta)$ is the coordinate of strut end or the mass center of body trunk with respect to the origin of body frame and $w^b(\theta)$ is the mass center of the whole system with respect to the origin of body frame. Applying the partial derivative of a position vector with respect to the orientation angle of the body frame (refer to equation (\ref{eqn:partial_psi})), we have
\begin{equation}
	\frac{\partial g_i(q)}{\partial q} = \begin{mat}{c}\displaystyle\frac{\partial g_i(q)}{\partial w} \\ \displaystyle\frac{\partial g_i(q)}{\partial\phi} \\ \displaystyle\frac{\partial g_i(q)}{\partial\theta} \end{mat} = \begin{mat}{c} \displaystyle I_3 \\ \displaystyle P(\phi)^\intercal Q(g_i^b(\theta)-w^b(\theta))\Omega(\phi) \\ \displaystyle\left(\frac{\partial g_i^b(\theta)}{\partial\theta} - \frac{\partial w^b(\theta)}{\partial\theta} \right)\Omega(\phi) \end{mat},
\end{equation}
where $P(\phi)\in\IR^{3\times3}$ is a matrix-valued function depending on the choice of the Euler angle sequence (see \ref{eqn:P_phi}), and for the 3-2-1 Euler angles it is given by (\ref{eqn:P_phi_321}), and the operator $Q(\cdot)$ is given by (\ref{eqn:Q_def}).
Given that the fluid force $f_i(q)=\Omega(\phi)^\intercal f_i^b(q)$, where $f_i^b(q)$ is the fluid force vector expressed in the body frame, the generalized force by fluid is given by
\begin{equation}\label{eqn:generalized_fluid_force}
	\varphi_f(q) 
	=\begin{mat}{c}\displaystyle\Omega(\phi)^\intercal \sum_{i=1}^n f_i^b(q) \\ \displaystyle P(\phi)^\intercal \sum_{i=1}^n Q(g_i^b(\theta)-w^b(\theta)) f_i^b(q) \\ \displaystyle \sum_{i=1}^n\left(\frac{\partial g_i^b(\theta)}{\partial\theta} - \frac{\partial w^b(\theta)}{\partial\theta} \right)f_i^b(q)\end{mat}.
\end{equation}
The fluid force $f_i^b(q)$ can be given from computational fluid dynamics or from the analytical fluid force model \cite{liu:16,taylor:52,chen:21}. In this study, the analytical fluid force model is used to provide quick and qualitative insights before going into the hydrodynamic details. The model for $f_i^b(q)$ is given in Appendix~\ref{appendix:fluid_force}, and $g_i^b(\theta)$, $w^b(\theta)$ are given in Appendix~\ref{appendix:generalized_force}.
Then, the generalized force in the equation of motion of the system (\ref{eqn:eom_theta}) and (\ref{eqn:eom_w}) is given by 
\begin{equation}
	\begin{mat}{c} h_w \\ \hdashline h_\vartheta\end{mat} = \begin{mat}{c}\mathbf{0} \\ \hdashline \mathbf{0} \\ \displaystyle\frac{\partial l(\theta)}{\partial\theta}\tau(\theta) \end{mat} + \varphi_f(q).
\end{equation}

\section{Simulation {\red examples}}\label{sec:results}
{\red We use a simple analytical fluid force model to test the dynamic model derived in the previous section. These simulations provide preliminary results, which can later be evaluated using a more accurate computational fluid dynamics (CFD) model.} We first simulate straight-line swimming to show what actuation signals are needed to generate the chord- and span-wise traveling waves on the fin. Then we analyze the influence of fin stiffness and resonance exploitation on swim speed (body length (BL) per second), stride length (the distance traveled per cycle normalized by the BL), propulsion efficiency (defined as $\eta = Tv/P_{in}$, where $T$, $v$, and $P_{in}$ are respectively the time-averaged thrust, velocity, and power input from actuation signals $u_i$ during steady-state swimming), and the ratio of energy loss due to fin damping to the total input energy from actuation signals $u_i$. Three values of fin stiffness are used, and for each value, these swimming performance quantities are plotted against the fin flapping frequency in the range around the fin resonance frequency. Finally, the somersault, rolling, and turning motions are simulated.

The fin is actuated as follows. Each elastic cable consists of a spring connected in series between two rigid cable segments. The model is actuated through the input $u$ in the second equation of (\ref{eqn:tau}), where motors roll in or release the rigid portions of the elastic cables, thereby changing the rest lengths $l_{i_0}$ of the elastic cables from their initial equilibrium values by $u_i$ (refer to (\ref{eqn:tau})). The signals $u_i$ ($i=1,\cdots,40$) are sinusoidal, and the amplitudes $|u_i|$ are given by $|u_i|=a_il_{i_0}$, where $a_i$ are set proportional to the smallest thickness of each tensegrity unit, with the proportionality constant equal to $1.5/\max(\mathrm{wing\;thickness})$, except for Fig.~\ref{fig:attitude} (somersault, rolling, and turning motions) where the proportionality constant is $0.5/\max(\mathrm{wing\;thickness})$. 
The cable damping ($c_i$ in (\ref{eqn:tau})) is added to make the first mode damping ratio 0.05 for resonance exploitation analysis (Fig.~\ref{fig:swimming_performance_total}) and 0.2 for all other simulations. The other model parameters are given in Appendix~\ref{sec:model_parameters}.

\subsection{Difference in speeds of actuation and fin kinematic waves}\label{sec:wing_kinematics}

\begin{figure}[!t]
	\centering
	\begin{minipage}{0.25\textwidth}
			\centerline{\epsfig{file=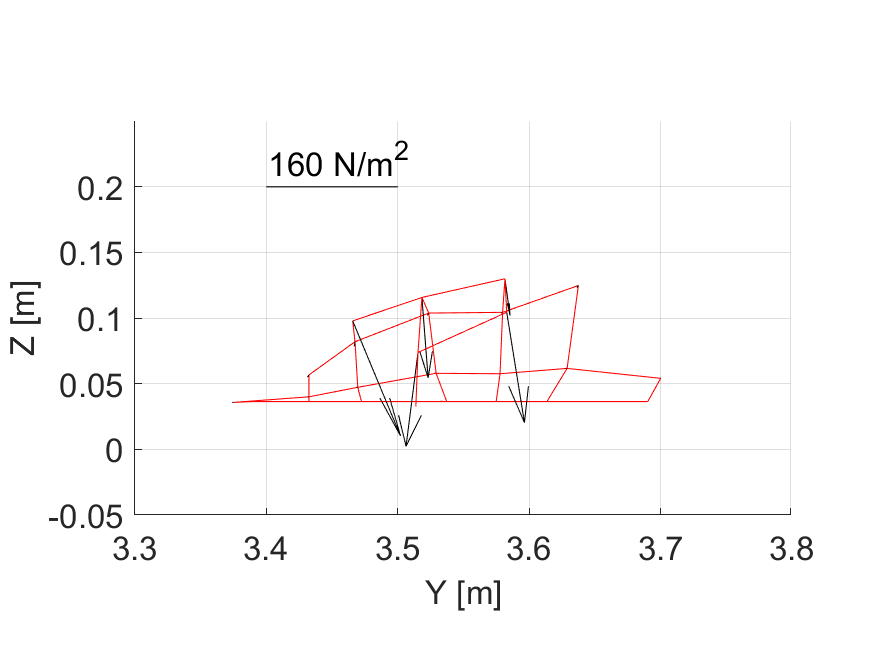,width=\textwidth}}
		\end{minipage}
		\begin{minipage}{0.25\textwidth}
			\centerline{\epsfig{file=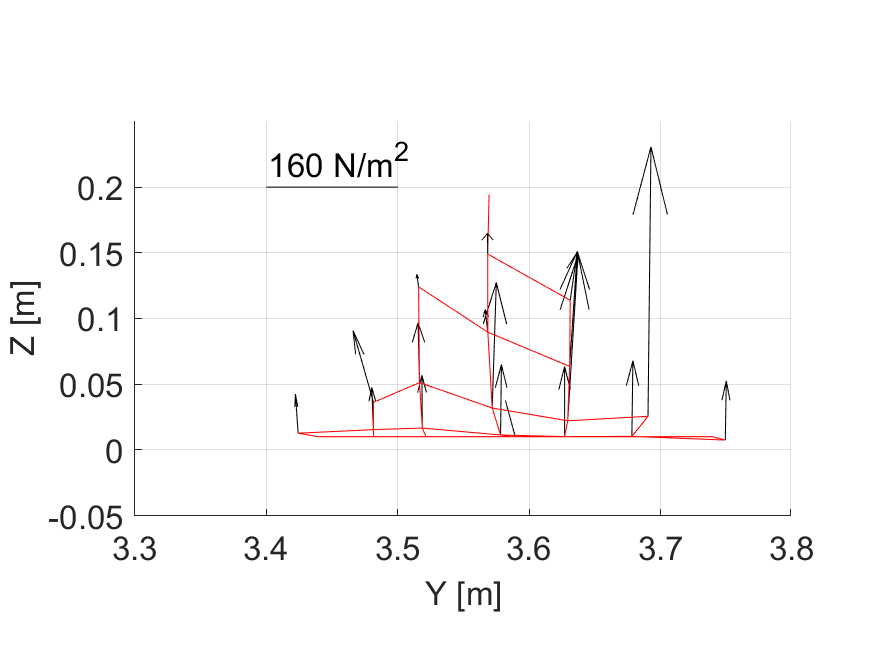,width=\textwidth}}
		\end{minipage}
		\begin{minipage}{0.25\textwidth}
			\centerline{\epsfig{file=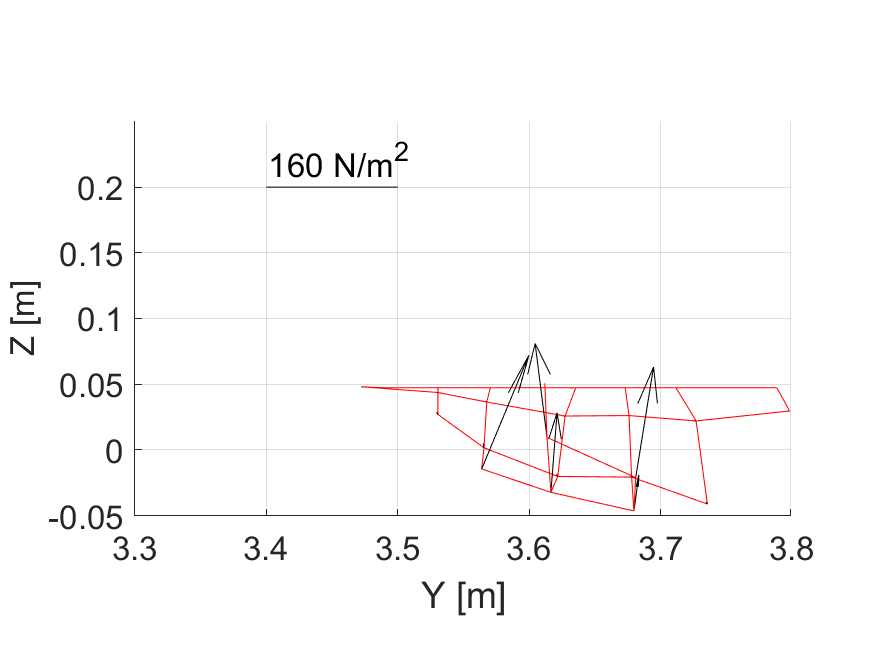,width=\textwidth}}
		\end{minipage}\\
		\begin{minipage}{\textwidth}
			\centering (a)
		\end{minipage}
		\begin{minipage}{0.25\textwidth}
			\centerline{\epsfig{file=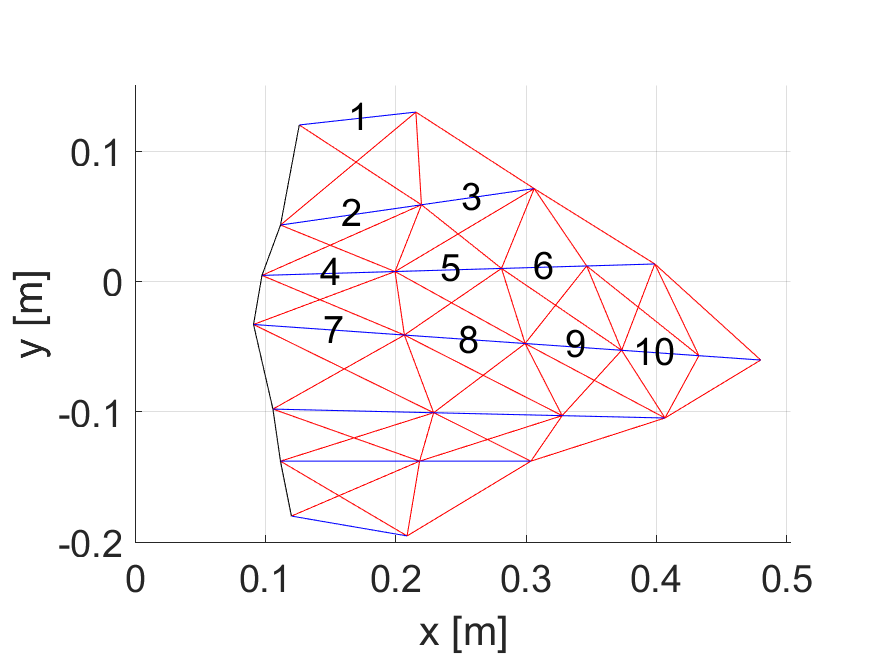,width=\textwidth}}
		\end{minipage}\hspace{5mm}
			\begin{minipage}{0.25\textwidth}
			\centerline{\epsfig{file=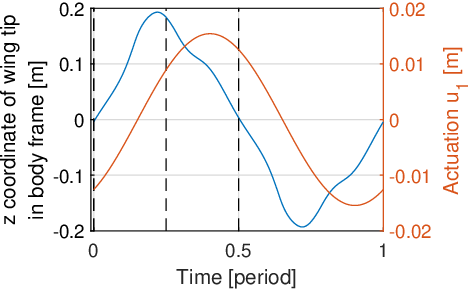,width=\textwidth}}
		\end{minipage}
		\\
		\begin{minipage}{\textwidth}
			\hspace{4.3cm} (b) \hspace{3.2cm} (c)
		\end{minipage}
		\begin{minipage}{0.4\textwidth}
			\centerline{\epsfig{file=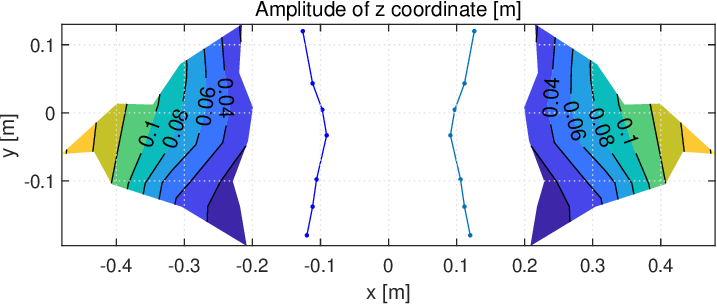,width=\textwidth}}
		\end{minipage}
		\begin{minipage}{0.4\textwidth}
			\centerline{\epsfig{file=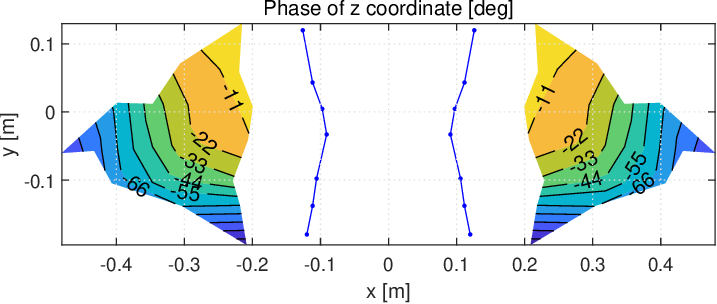,width=\textwidth}}
		\end{minipage}
		\begin{minipage}{\textwidth}
			\centering (d)
		\end{minipage}
		\caption{(a) Kinematics of the virtual fin surface (red lines) and fluid force vectors (black arrows) during free swimming, where all actuated tensegrity units in the first four radials are actuated in the same phase. Three snapshots correspond to the time instants within one cycle indicated by the vertical dashed lines in (b). 
		(b) $z$-direction displacement of the fin tip over one cycle in the body frame. (c) The actuation signal $u_1$ for the cable on the top surface of the first tensegrity radial. (d) Amplitude and phase of the $z$-direction displacement of the fin surface. Fin displacement is measured relative to the mass center of the whole system. The blue lines mark the places where the fins are attached to the body trunk.}\label{fig:flap_total}
\end{figure}

Fig.~\ref{fig:flap_total}a shows the kinematics of the virtual fin surface (the average displacement of the fin top and bottom surfaces) and the fluid force vectors during free straight-line swimming, where the actuated tensegrity units are numbered in Fig.~\ref{fig:flap_total}b and are all actuated simultaneously. Despite all tensegrity units being actuated in the same phase, both chord-wise and span-wise traveling waves emerge on the fin surfaces, as also indicated by the phase contour of the fin $z$-direction displacement in the body frame (Fig.~\ref{fig:flap_total}d). The fin exhibits a pitch motion, and the fin tip lags behind the fin surface points closer to the body trunk. The three snapshots in Fig.~\ref{fig:flap_total}a correspond respectively to the time instants when the fin tip passes upward through the $x$-$y$ plane of the body frame, reaches the peak position, and passes downward through the $x$-$y$ plane (Fig.~\ref{fig:flap_total}c, left $y$-axis). The actuation signal to the cable on the top surface of the first actuated tensegrity unit is plotted on the right $y$-axis of Fig.~\ref{fig:flap_total}c; all other tensegrity units share the same actuation phase.
Fig.~\ref{fig:flap_total}d shows the phase and amplitude of the fin $z$-direction displacement measured in the body frame. The results indicate that the spatial wave number of the observed fin kinematics is larger than that of the actuation signals.

\subsection{Influence of fin stiffness and resonance exploitation on swimming speed}\label{sec:simulation_resonance}

\begin{figure}[!t]
	\centering
	\begin{minipage}{0.45\textwidth}
		\centerline{\epsfig{file=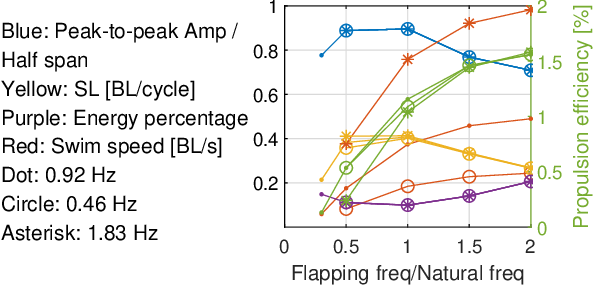,width=\textwidth}}
	\end{minipage}
	\hspace{5mm}
	\begin{minipage}{0.27\textwidth}
		\centerline{\epsfig{file=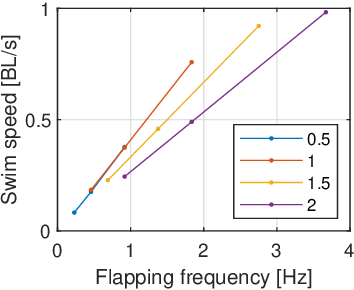,width=\textwidth}}
	\end{minipage}
	\begin{minipage}{\textwidth}
		\hspace{5.2cm} (a) \hspace{4cm} (b)
	\end{minipage}
	\caption{(a) Swimming performance for different fin stiffnesses. The markers distinguish fin stiffness cases with first-mode natural frequencies of 0.92~Hz (dot), 0.46~Hz (circle), and 1.83~Hz (asterisk). The colors represent: normalized peak-to-peak fin tip amplitude by the span from the body midline to fin tip (blue), swim speed (BL/s) (red), stride length (BL/cycle) (yellow), and the ratio of cable damping energy loss to the total input energy (purple). (b) Swim speed increases linearly with fin flapping frequency when resonance is exploited. Three data points in each line represent the fin natural frequency varying from 0.46~Hz to 0.92~Hz and 1.83~Hz, and the line color indicates the fin flaps at 0.5, 1, 1.5, or 2 times its natural frequency.}\label{fig:swimming_performance_total}
\end{figure}

Resonance exploitation of the body-fluid system is a control mechanism observed in biological swimmers. To study its effect, the spring constants of elastic cables in a single fin model are varied to change the fin's natural frequency, with the fin root fixed to the inertial frame and no fluid force applied. The first-mode natural frequencies of the fin, computed from a linearized fin model, are 0.92~Hz, 0.46~Hz, and 1.83~Hz respectively.
The actuation pattern and amplitudes of $u_i$ are the same as in the previous section, i.e., all actuated tensegrity units are actuated simultaneously, except that the fin flapping frequency now varies from 0.5 to 2 times the fin natural frequency. Fig.~\ref{fig:swimming_performance_total}a shows that when the fins flap at the natural frequency, the peak-to-peak amplitude of the fin tip (blue lines) and the stride length (yellow lines) reach their maximum values, while the percentage of energy loss due to cable damping in the total input energy from actuation signals $u_i$ is minimized (purple lines). The stride length, fin flapping amplitude, and the percentage of energy loss due to cable damping are nearly independent of fin stiffness when fin resonance is exploited. These results suggest that, for a given swim speed whose flapping frequency is determined by hydrodynamic resonance \cite{kohannim:14,triantafyllou:93,gopalkrishnan:94,anderson:98}, fin stiffness should be tuned to align the fin natural frequency with the flapping frequency, so as to maximize fin flapping amplitude and stride length while minimizing the fraction of energy lost to fin damping. This is consistent with the biological observation that animals adjust body stiffness at different swimming speeds \cite{kohannim:14,schwalbe:18,mchenry:95,long:96,long:98,tytell:16}. Fig.~\ref{fig:swimming_performance_total}a also shows that swim speed (BL/s, red lines) increases with both fin stiffness and flapping frequency.
Fig.~\ref{fig:swimming_performance_total}b re-plots the relationship between swim speed and fin flapping frequency for the data points in Fig.~\ref{fig:swimming_performance_total}a. It shows that swim speed increases linearly with flapping frequency when fin stiffness is tuned so that the natural frequency matches the flapping frequency, but this linearity is lost when fin stiffness is fixed. The linear relationship between swim speed and flapping frequency has been observed in manta ray swimming and reproduced in simulations coupling a potential flow model with morphing wing boundary surfaces \cite{fish:16}. Our results therefore suggest that manta rays stiffen their fins to swim faster.
The propulsion efficiency $\eta$ (green lines in Fig.~\ref{fig:swimming_performance_total}a) is very low, as the fluid force component perpendicular to the fin surface is much larger than the average thrust.

\subsection{Effects of body inertial motion on fin kinematics}\label{sec:simulation_outOfPhase}

\begin{figure}[!t]
	\centering
	\begin{minipage}{0.25\textwidth}
			\centerline{\epsfig{file=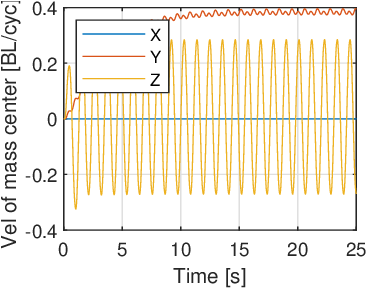,width=\textwidth}}
		\end{minipage}	\hs
	\begin{minipage}{0.25\textwidth}
			\centerline{\epsfig{file=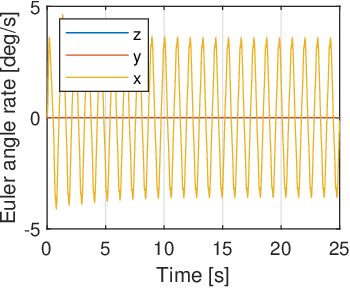,width=\textwidth}}
		\end{minipage}
	\begin{minipage}{0.35\textwidth}
			\centerline{\epsfig{file=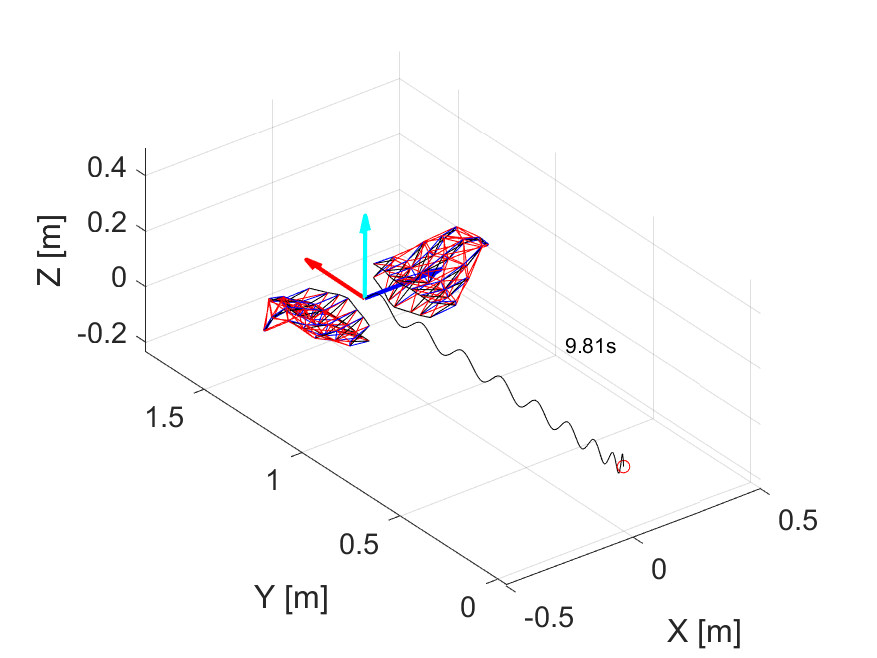,width=\textwidth}}
		\end{minipage}\\
	\begin{minipage}{\textwidth}
	\hspace{2.2cm} (a) \hspace{3.1cm} (b) \hspace{3.4cm} (c)
	\end{minipage}
    \begin{minipage}{0.25\textwidth}
			\centerline{\epsfig{file=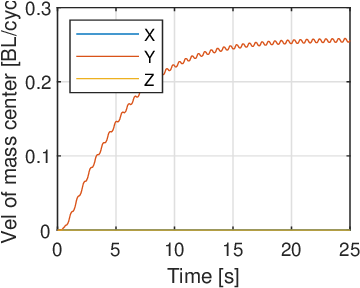,width=\textwidth}}
		\end{minipage}\hs
	\begin{minipage}{0.25\textwidth}
			\centerline{\epsfig{file=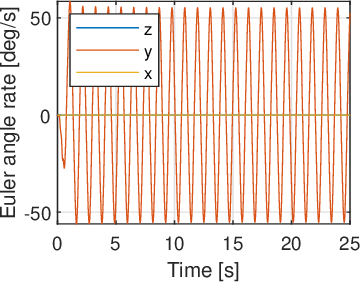,width=\textwidth}}
		\end{minipage}
	\begin{minipage}{0.35\textwidth}
			\centerline{\epsfig{file=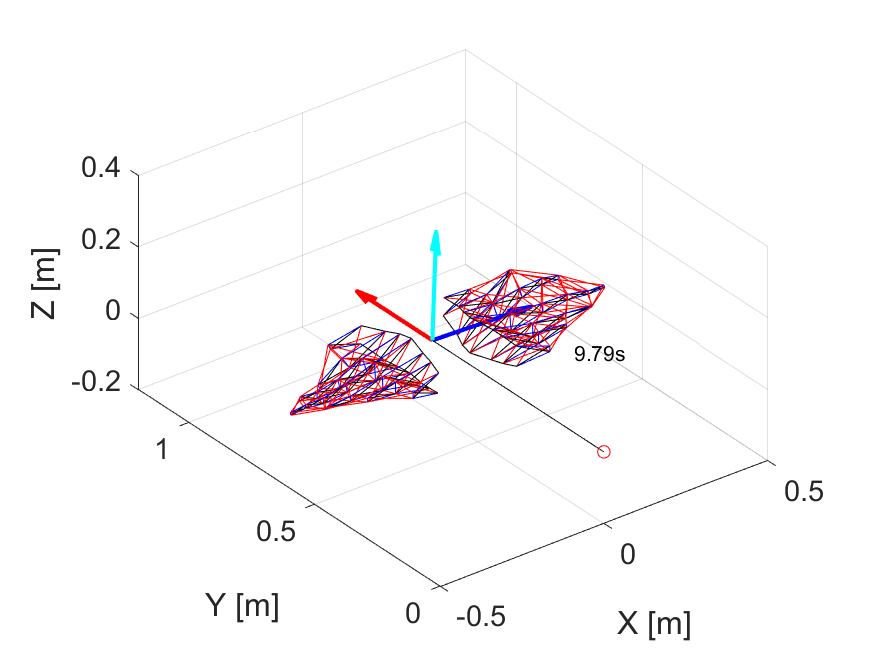,width=\textwidth}}
		\end{minipage}\\
		\begin{minipage}{\textwidth}
			\hspace{2.2cm} (d) \hspace{3.1cm} (e) \hspace{3.4cm} (f)
		\end{minipage}
	\begin{minipage}{0.4\textwidth}
			\centerline{\epsfig{file=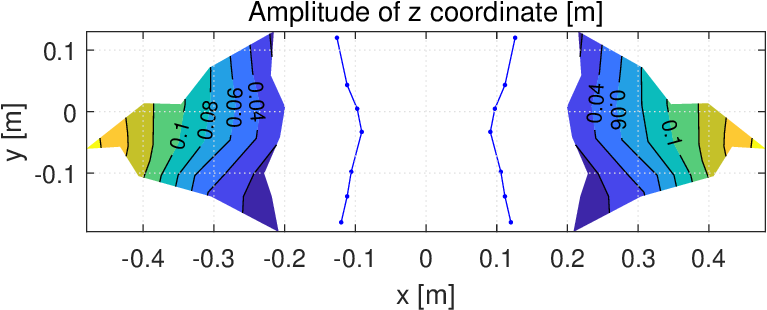,width=\textwidth}}
		\end{minipage}
	\begin{minipage}{0.4\textwidth}
			\centerline{\epsfig{file=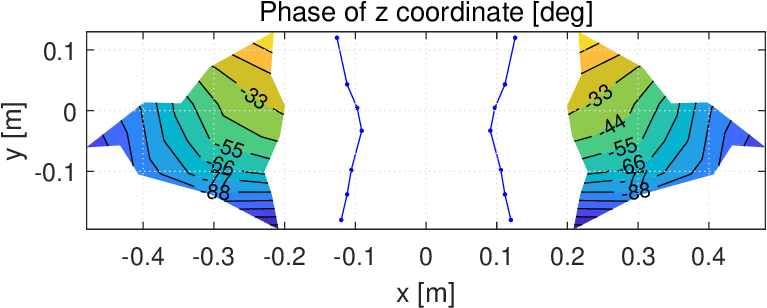,width=\textwidth}}
		\end{minipage}\label{fig:in_phase_phase} \\
	\begin{minipage}{\textwidth}
			\hspace{3.8cm} (g) \hspace{4.7cm} (h)
	\end{minipage}
	\begin{minipage}{0.4\textwidth}
			\centerline{\epsfig{file=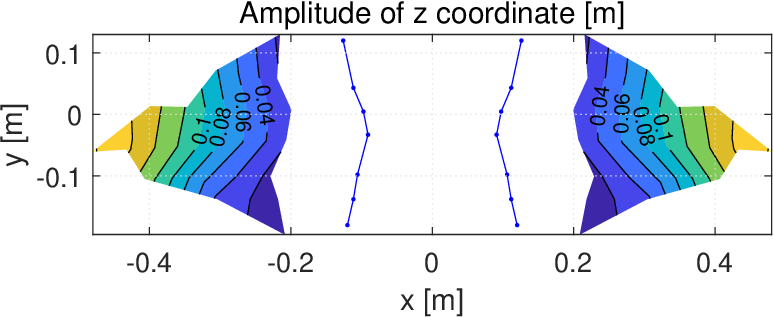,width=\textwidth}}
		\end{minipage}
	\begin{minipage}{0.4\textwidth}
			\centerline{\epsfig{file=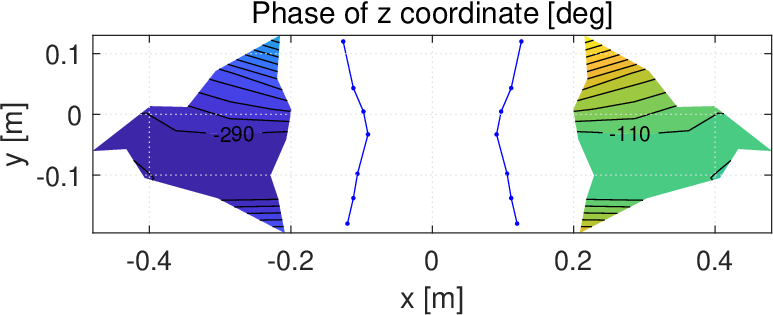,width=\textwidth}}
		\end{minipage}
	\begin{minipage}{\textwidth}
			\hspace{3.8cm} (i) \hspace{4.7cm} (j)
	\end{minipage}
	\caption{Velocity of the mass center of the whole system (a,d), rates of the Euler angles of the body trunk (b,e), trajectory of the origin of the body frame (c,f), and amplitude (g,i) and phase (h,j) of the $z$-direction displacement of fin surfaces in the body frame, for in-phase (a--c, g--h) and out-of-phase (d--f, i--j) fin flapping modes. The phase spacing in (j) is $10^o$.}\label{fig:out_of_phase_total}
\end{figure}

Both in-phase and out-of-phase flapping of the two fins are observed in butterfly rays. We simulate these two swimming modes. {\red The results show that the fin kinematic wave can travel faster than tension actuation wave in out-of-phase flapping straight-line swimming due to the body inertial roll motion and the resulting fluid forces.}
%
In the simulation of this subsection, the actuation within the tensegrity radials remains the same, but the intersegmental phase lag between the radials for $u_i$ is increased from $0^o$ (the value used in Fig.~\ref{fig:flap_total}) to $12^o$ for the in-phase flapping mode and $54^o$ for the out-of-phase flapping mode, each tuned to maximize the swim speed of the respective case. The amplitudes of $u_i$ remain the same.
The results show that in-phase flapping produces a combined pitch and heave motion of the body frame, while out-of-phase flapping produces a roll motion. The swim speed is higher in the in-phase flapping mode (Fig.~\ref{fig:out_of_phase_total}).
{\red Fig.~\ref{fig:out_of_phase_total}j shows that the roll motion of the body frame causes the fin kinematic wave speed to be faster than the actuation wave $u_i$ (actuation phase lag of the first four actuated tensegrity radials: $162^o$). This contrasts with the in-phase flapping case, where the fin kinematic wave is slightly slower than the actuation wave $u_i$ (Fig.~\ref{fig:out_of_phase_total}h, actuation phase lag: $36^o$). In the out-of-phase case, the phases in the posterior region of the fin advance those in the middle portion (Fig.~\ref{fig:out_of_phase_total}j). This is because the roll motion causes fluid forces that simultaneously bend the short tensegrity radials in the posterior region together with those in the anterior region. In contrast, in the in-phase flapping mode, the heave and pitch motions have an opposing effect on the fin surface normal velocities and the resulting fluid forces, so the phase advance in the posterior radials is not observed.}

\subsection{Somersault, rolling and turning via fin kinematics}\label{sec:simulation_attitude}

\begin{figure}[!t]
	\centering
	\begin{minipage}{0.35\textwidth}
			\includegraphics[width=\textwidth,trim=40mm 140mm 110mm 90mm,clip=true]{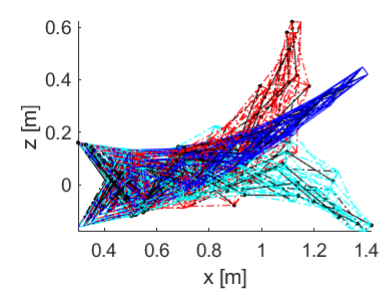}
			\centering (a)
		\end{minipage}
	\begin{minipage}{0.5\textwidth}
			\includegraphics[width=\textwidth,trim=20mm 0mm 20mm 10mm,clip=true]{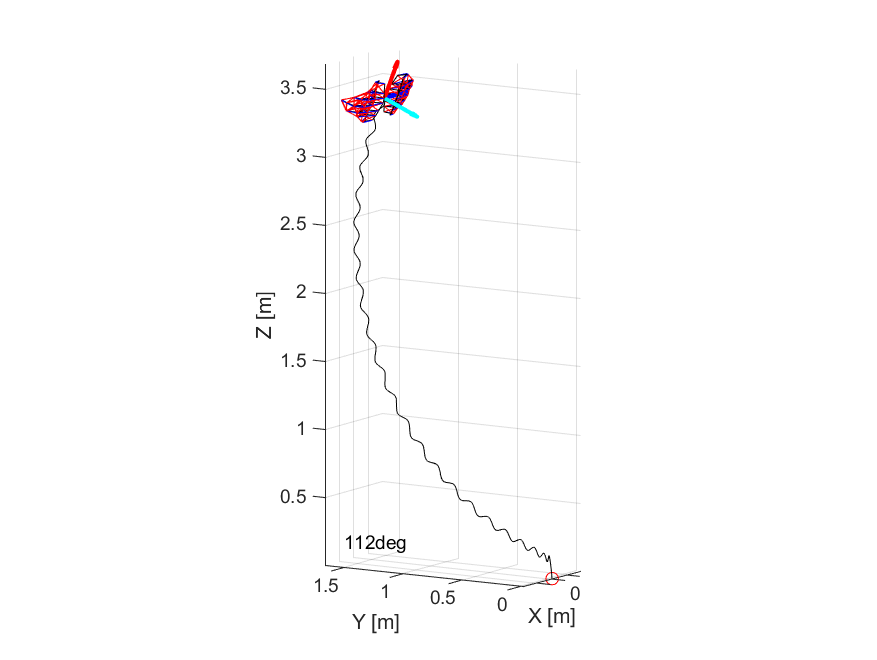}
			\centering (b)
		\end{minipage}\hfil
	\begin{minipage}{0.4\textwidth}
			\includegraphics[width=\textwidth,trim=20mm 0mm 10mm 10mm,clip=true]{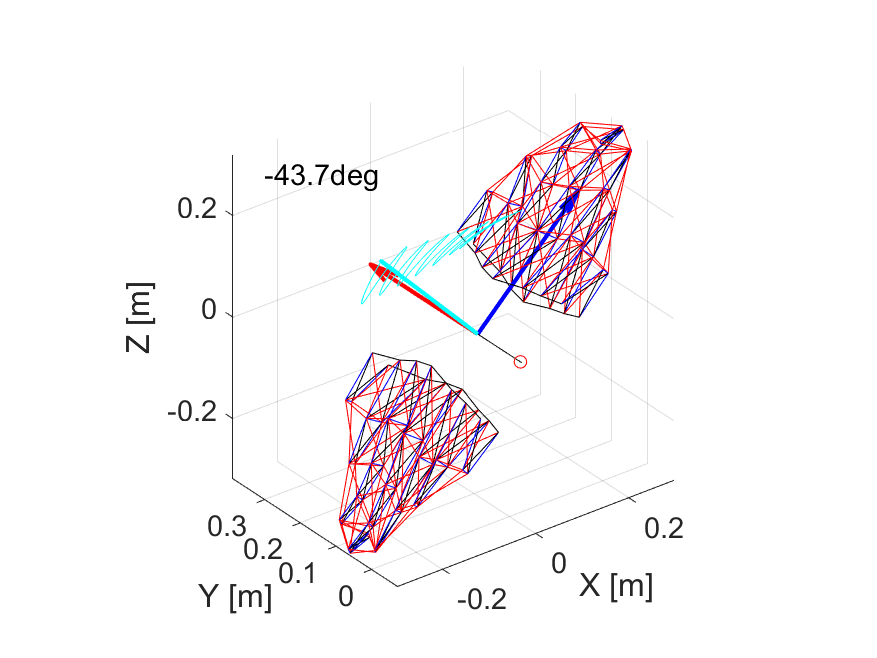}
			\centering (c)
		\end{minipage}
	\begin{minipage}{0.4\textwidth}
			\includegraphics[width=\textwidth,trim=0mm 0mm 10mm 10mm,clip=true]{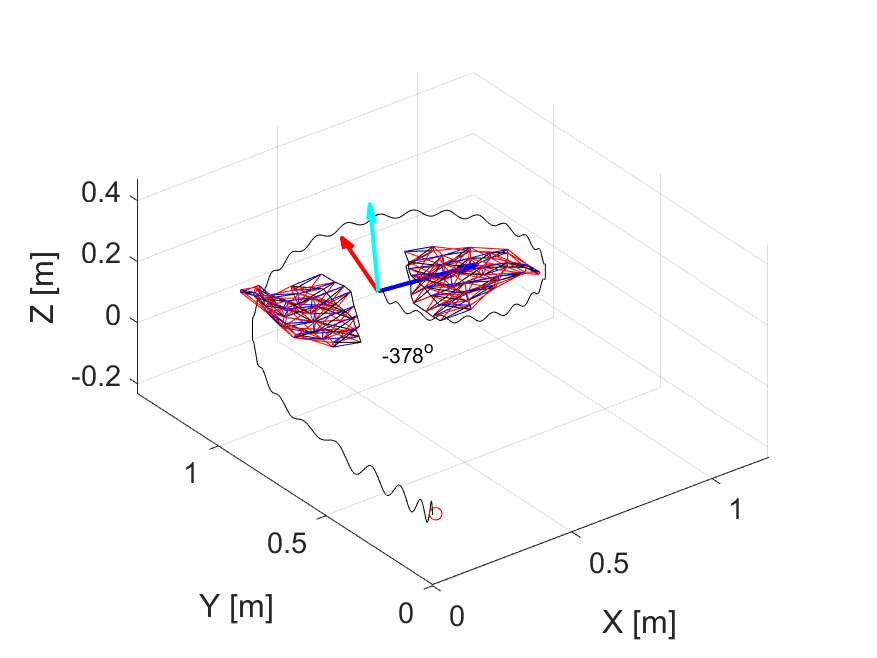}
			\centering (d)
		\end{minipage}
	\caption{(a) The fin first deforms into a curved configuration (blue-colored snapshot) and then flaps about it between two extreme positions (red- and cyan-colored snapshots), generating net pitch and roll torques. Adapted from \cite{chen:21}. (b,c,d) The somersault, rolling, and turning motions. The black curves are the trajectories of the mass center of the whole system, and the cyan curve in (c) shows the trajectory of a point on the $z$-axis of the body frame. During turning, the actuation to the right fin is set to zero and the right fin oscillates passively under the fluid force.}\label{fig:attitude}
\end{figure}

We demonstrate how to use the two fins to generate somersault, rolling, and turning motions. When a fin deforms into a curved configuration (Fig.~\ref{fig:attitude}a) and then flaps about that configuration, net pitch and roll torques are produced. The somersault motion in Fig.~\ref{fig:attitude}b is generated when both fins deform in the same direction into the blue-colored configuration in Fig.~\ref{fig:attitude}a and then flap about it in phase, producing a net pitch torque. The rolling motion in Fig.~\ref{fig:attitude}c is generated when the two fins deform in opposite directions into the blue-colored configuration and then flap about it out of phase, producing a net roll torque. The roll angle exhibits an oscillatory back-and-forth motion as seen in Fig.~\ref{fig:attitude}c. For the turning maneuver in Fig.~\ref{fig:attitude}d, starting from a swim speed of 0.16~BL/s, the actuation signals to the right fin are set to zero. The turning radius is positively correlated with swim speed, and decreases as the turning rate grows with continued flapping of the left fin.

\section{Conclusion}\label{sec:conclusion}

We modeled the active deformation of biological pectoral fins using a tensegrity structure and developed a body-fluid interaction model to simulate batoid swimming. The model is exercised through simulation examples using an analytical fluid force model, which, while simplified, is sufficient to reveal qualitative trends in the system behavior. The simulations suggest the following trends. The locomotion quantities (stride length, fin flapping amplitude, and the percentage of energy loss due to fin damping in the total input energy) appear invariant with respect to swim speed when fin resonance is exploited. The linear relationship between swim speed and flapping frequency, which is observed in manta ray swimming, appears to be lost when resonance is not exploited, suggesting that the pectoral fin is stiffened for faster swimming. Body inertial motion, in addition to fin stiffness, also influences the fin kinematic waves. Simulations of in-phase and out-of-phase flapping indicate that the wave speed of fin kinematics can be either faster or slower than the tension actuation wave. These qualitative trends should be further evaluated and quantified through simulations that replace the analytical fluid force model with a computational fluid dynamics model. Such a coupled model may be used to generate richer hydrodynamic information during swimming or maneuvering, or to predict neural control signals when muscle dynamics are incorporated.










\printbibliography

\renewcommand{\thetable}{A\arabic{table}}
\setcounter{table}{0}
\renewcommand{\thefigure}{A\arabic{figure}}
\setcounter{figure}{0}
\renewcommand{\theequation}{A-\arabic{equation}}
\setcounter{equation}{0}
\renewcommand{\thesection}{A\arabic{section}}
\setcounter{section}{0}
\section*{Appendix}
	
	\section{General equation of motion of multiple rigid body system}\label{appendix:general_eom}
	The total kinetic energy of a system of $n$ rigid bodies (body trunk and struts) that are connected in an arbitrary manner is
	\begin{equation}\label{eqn:energy_multbody}
		T=\frac{1}{2}\sum_{i=1}^n (m_i||\dot r_i||^2+\omega_i^\intercal J_i\omega_i),
	\end{equation}
	where the subscript $i$ denotes the $i^{th}$ rigid body, $r_i$ is the position vector for the mass center expressed in the inertial frame, $m_i$ is the mass, $J_i$ is the moment of inertia around the mass center and $\omega_i$ is the angular velocity vector in the body frame. The position of the mass center of the $i^{th}$ rigid body $r_i$ is given by $r_i = w+p_i(\vartheta)$ for some function $p_i$, where $w\in\IR^3$ is the position of the mass center of the whole system, and
	\begin{equation}
		\dot r_i = \dot w +\left(\frac{\partial p_i}{\partial\vartheta}\right)^\intercal\dot\vartheta
	\end{equation}
	The kinetic energy is then expressed in terms of the position of the mass center of the whole system in the inertial frame $w\in\IR^3$, the orientation of the body frame $\phi\in\IR^3$, and the body shape variables $\theta\in\IR^{n-1}$ as
	\begin{equation}\label{eqn:J_vartheta_tmp}
		T=\frac{1}{2}m_o||\dot w||^2 + \frac{1}{2}\dot\vartheta^\intercal \mathbb{J}(\vartheta)\dot\vartheta,
	\end{equation}
	where $\vartheta=\begin{bmatrix}\phi^\intercal & \theta^\intercal\end{bmatrix}^\intercal$, $m_o=\sum_{i=1}^n m_i$, $\mathbb{J}(\vartheta)$ is given by (\ref{eqn:J_vartheta_o}),
	and we note that
	\[
	\sum_{i=1}^n m_i\dot{w}^\intercal\left(\frac{\partial p_i}{\partial\vartheta}\right)^\intercal\dot\vartheta = \dot{w}^\intercal\left(\frac{\partial}{\partial\vartheta}\left(\sum_{i=1}^n m_ip_i(\vartheta)\right)\right)^\intercal\dot\vartheta = 0
	\]
	holds because $w$ is taken as the mass center of the system and therefore $\sum_{i=1}^nm_ip_i(\vartheta)=0$.
	
	
	The Euler-Lagrange equation is given by
	\[
	\frac{d}{dt}\left(\frac{\partial L}{\partial\dot q}\right) - \frac{\partial L}{\partial q} = h
	\]
	where $L=T-V$ is the Lagrangian and $h$ is the generalized force. If we take the elastic cable forces in tensegrity radials as the external forces, and therefore $V=0$, the Euler-Lagrange equation reduces to (\ref{eqn:eom_theta}), (\ref{eqn:eom_w}) with (\ref{eqn:J_vartheta_o}) and (\ref{eqn:G_vartheta}).

	\section{Body dynamics}\label{appendix:body_dynamics}
	The matrix form expressions for $\mathbb{J}(\vartheta)$ and $G(\vartheta,\dot\vartheta)$ in (\ref{eqn:J_vartheta_o}) and (\ref{eqn:G_vartheta}) are derived for the batoid swimming. 
	Let the origin of body frame expressed in the inertial frame be denoted by $w_o$ and $p_i^b(\theta)$ denote the vector from the origin of body frame to the mass center of the $i^{th}$ rigid body expressed in the body frame. The position of the mass center of the $i^{th}$ rigid body expressed in the inertial frame and the mass center of the whole system $w$ are respectively given by
	\[
	r_i=w_o+\Omega(\phi)^\intercal p_i^b(\theta),  \hs w=\frac{1}{m_o}\sum_{i=1}^n m_ir_i=w_o+\frac{1}{m_o}\Omega(\phi)^\intercal\sum_{i=1}^n m_ip_i^b,
	\]
	where the superscript $b$ in $p_i^b$ represents the coordinates in body frame, $\phi\in\IR^3$ is the Euler angle vector that specifies the orientation of the body frame, which is fixed on the body trunk, $\theta = \begin{bmatrix} \theta_1 & \cdots & \theta_{n-1} \end{bmatrix}^\intercal$ is the vector of strut angles with respect to the horizontal plane (refer to Fig.~\ref{fig:config_whole_body}b), and $n$ is the total number of components, including the rigid body trunk and the struts of two wings. Then, $r_i$ is expressed in the form of (\ref{eqn:ri}) in terms of the mass center of the whole system as
	\begin{equation}\label{eqn:pi}
		r_i= w + p_i(\vartheta), \hs p_i(\vartheta)=\Omega(\phi)^\intercal (p_i^b(\theta)-w^b(\theta)), \hs \vartheta=\begin{bmatrix}\phi \\ \theta \end{bmatrix}, \hs w^b(\theta):= \frac{1}{m_o}\sum_{i=1}^n m_ip_i^b(\theta),
	\end{equation}
	where $w^b(\theta)$ is the mass center of the whole system coordinatized in the body frame.
	
	The angular velocity of the strut $i$, $\omega_i$ ($i=1,\cdots,n-1$), is given by the sum of angular velocity of the body frame $P(\phi)\dot\phi$ and the angular velocity of the strut relative to the body frame $e_2\dot\theta_i$, where $e_2 =\begin{bmatrix} 0 & 1 & 0\end{bmatrix}^\intercal$. The angular velocity $\omega_i$ expressed in the frame fixed on the mass center of the strut is given by:
	\begin{equation}\label{eqn:omegai}
		\omega_i = \Omega_{\theta_i}\Omega_{\zeta_i} P(\phi)\dot\phi + \dot\theta_i e_2 = \Omega_{\theta_i}\Omega_{\zeta_i} P(\phi)\dot\phi + e_2e_i^\intercal\dot\theta,
	\end{equation}
	where $\Omega_{\theta_i}$ and $\Omega_{\zeta_i}$ transform the angular velocity of the body frame to the frame that is fixed on the strut mass center,
	\[
	\Omega_{\theta_i}=\begin{bmatrix} \cos\theta_i & 0 & -\sin\theta_i \\ 0 & 1 & 0 \\ \sin\theta_i & 0 & \cos\theta_i \end{bmatrix}, \hs \Omega_{\zeta_i} = \begin{bmatrix}\cos\zeta_i & \sin\zeta_i & 0 \\ -\sin\zeta_i & \cos\zeta_i & 0 \\ 0 & 0 & 1 \end{bmatrix},
	\]
	$\zeta_i$ and $\theta_i$ ($i=1,\cdots, n-1$) are the azimuth and elevation angles of strut $i$, $e_i\in\IR^{n-1}$ is a vector with the $i^{th}$ element being 1 and the rest being 0's. The angle velocity of the body trunk is the same as the body frame, i.e., $P(\phi)\dot\phi$. Then, $P_i(\vartheta)$ in (\ref{eqn:J_vartheta_o}) is given by
	\begin{equation}\label{eqn:P_i}
		P_i(\vartheta) = \begin{bmatrix}\Omega_{\theta_i}\Omega_{\zeta_i}P(\phi) & e_2e_i^\intercal\end{bmatrix} \hs P_i(\vartheta)=\begin{bmatrix}P(\phi) & \bf{0} \end{bmatrix}
	\end{equation}
	respectively for the struts ($i=1,\cdots,n-1$) and for the body trunk ($i=n$).
	
	Plug (\ref{eqn:pi}) for the expression of $p_i(\vartheta)$ and (\ref{eqn:P_i}) for $P_i(\vartheta)$ into (\ref{eqn:J_vartheta_o}) and (\ref{eqn:G_vartheta}) to evaluate $\mathbb{J}(\vartheta)$ and $G(\vartheta,\dot\vartheta)$ in the equation of motion (\ref{eqn:eom_theta}), which are evaluated next.
	According to (\ref{eqn:partial_psi}), 
	\[
	\left(\frac{\partial p_i}{\partial\phi}\right)^\intercal = \Omega(\phi)^\intercal Q(p_i^b-w^b)^\intercal P(\phi).
	\]
	Then, the matrix $\mathbb{J}(\vartheta)$ becomes
	\begin{align}\label{eqn:J_vartheta}
		\mathbb{J}(\vartheta) = &\sum_{i=1}^n m_i\left(\frac{\partial p_i}{\partial\vartheta}\right)\left(\frac{\partial p_i}{\partial\vartheta}\right)^\intercal + \left(\frac{\partial\omega_i}{\partial\dot\vartheta}\right) J_i\left(\frac{\partial\omega_i}{\partial\dot\vartheta}\right)^\intercal \cr
		= &\sum_{i=1}^n m_i\begin{bmatrix}P(\phi)^\intercal Q(p_i^b - w^b)Q(p_i^b-w^b)^\intercal P(\phi) & P(\phi)^\intercal Q(p_i^b - w^b)\displaystyle \left(\frac{\partial p_i^b}{\partial\theta} - \frac{\partial w^b}{\partial\theta}\right)^\intercal \\ \displaystyle\left(\frac{\partial p_i^b}{\partial\theta} - \frac{\partial w^b}{\partial\theta}\right)Q(p_i^b-w^b)^\intercal P(\phi) & \displaystyle\left(\frac{\partial p_i^b}{\partial\theta} - \frac{\partial w^b}{\partial\theta}\right)\left(\frac{\partial p_i^b}{\partial\theta} - \frac{\partial w^b}{\partial\theta}\right)^\intercal \end{bmatrix} \cr
		&+\sum_{i=1}^{n-1}\begin{bmatrix} P(\phi)^\intercal\Omega_{\zeta_i}^\intercal\Omega_{\theta_i}^\intercal J_i \Omega_{\theta_i}\Omega_{\zeta_i}P(\phi)  & P(\phi)^\intercal\Omega_{\zeta_i}^\intercal\Omega_{\theta_i}^\intercal J_i e_2e_i^\intercal \\ e_ie_2^\intercal J_i \Omega_{\theta_i}\Omega_{\zeta_i}P(\phi) & e_ie_2^\intercal J_i e_2e_i^\intercal \end{bmatrix} + \begin{bmatrix}P(\phi)^\intercal J_bP(\phi) & 0 \\ 0 & 0 \end{bmatrix} \cr
		=& \begin{bmatrix} P(\phi)^\intercal J(\theta)P(\phi) & P(\phi)^\intercal\eta(\theta) \\ \eta(\theta)^\intercal P(\phi) & \sigma(\theta) \end{bmatrix}
	\end{align}
	where 
	\begin{align}
		J(\theta) =& \sum_{i=1}^n m_iQ(p_i^b - w^b)Q(p_i^b-w^b)^\intercal + \sum_{i=1}^{n-1} \Omega_{\zeta_i}^\intercal\Omega_{\theta_i}^\intercal J_i \Omega_{\theta_i}\Omega_{\zeta_i} + J_b \cr
		= & Q(p^b)(M\otimes I_3)Q(p^b)^\intercal + \Omega_\zeta^\intercal \Omega_\theta^\intercal J_o\Omega_\theta\Omega_\zeta + J_b \label{eqn:Jtheta} \\
		\eta(\theta) =& \sum_{i=1}^n m_i Q(p_i^b - w^b)\displaystyle \left(\frac{\partial p_i^b}{\partial\theta} - \frac{\partial w^b}{\partial\theta}\right)^\intercal + \sum_{i=1}^{n-1} \Omega_{\zeta_i}^\intercal\Omega_{\theta_i}^\intercal J_i e_2e_i^\intercal \cr
		=&  (y^\intercal MLC_\theta + z^\intercal MLS_\zeta S_\theta - \sin\zeta^\intercal J_2)\otimes e_1 \cr &- (x^\intercal MLC_\theta + z^\intercal MLC_\zeta S_\theta -\cos\zeta^\intercal J_2)\otimes e_2 \cr &
		+ (y^\intercal MLC_\zeta S_\theta - x^\intercal MLS_\zeta S_\theta)\otimes e_3 \label{eqn:eta_theta}\\
		\sigma(\theta) =& \sum_{i=1}^n m_i\left(\frac{\partial p_i^b}{\partial\theta} - \frac{\partial w^b}{\partial\theta}\right)\left(\frac{\partial p_i^b}{\partial\theta} - \frac{\partial w^b}{\partial\theta}\right)^\intercal + \sum_{i=1}^{n-1}  e_ie_2^\intercal J_i e_2e_i^\intercal \cr
		=& S_\theta P_\zeta S_\theta + C_\theta Q C_\theta + J_2, \hs Q=L^\intercal ML, \hs P_\zeta = C_\zeta QC_\zeta + S_\zeta Q S_\zeta, \label{eqn:sigma_theta}
	\end{align}
	and
	\begin{equation}\label{eqn:Qpb}
		Q(p^b)= z^\intercal\otimes E_3 + y^\intercal\otimes E_2 + x^\intercal\otimes E_1, \hs M=\mathrm{diag}(m_1,\cdots,m_n)-\frac{mm^\intercal}{m_o},
	\end{equation}
	\[
	E_1 =\begin{mat}{ccc} 0 & -1 & 0 \\ 1 & 0 & 0 \\ 0 & 0 & 1 \end{mat}, \hs E_2 =  \begin{mat}{ccc} 0 & 0 & 1 \\ 0 & 0 & 0 \\ -1 & 0 & 0 \end{mat}, \hs E_3 = \begin{mat}{ccc} 0 & 0 & 0 \\ 0 & 0 & -1 \\ 0 & 1 & 0 \end{mat}, \hs m=\begin{mat}{c} m_1 \\ \vdots \\ m_n \end{mat},
	\]
	\[
	p^b = x\otimes e_1 + y\otimes e_2 + z\otimes e_3, \hs e_1=\begin{bmatrix} 1 \\ 0 \\ 0\end{bmatrix}, \hs  e_2=\begin{bmatrix}0 \\ 1 \\ 0\end{bmatrix} \hs e_3=\begin{bmatrix} 0 \\ 0 \\ 1\end{bmatrix},
	\]
	and $x$, $y$ and $z$ are vectors of coordinates for the mass center of tensegrity struts and body trunk which are functions of strut angle vector $\theta$ and given by \cite{futakata:sice10}:
	\begin{equation}\label{eqn:xyz}
		x=LC_\zeta \cos\theta + x_0, \hs y= LS_\zeta \cos\theta +y_0, \hs z=L\sin\theta + z0,
	\end{equation}
	where
	\[
	L = \begin{bmatrix} \mathsf{L}_1 &  &  \\  & \ddots &  \\  &  & \mathsf{L}_N \\ 0 & \cdots & 0 \end{bmatrix}, \hs 
	\mathsf{L}_i = I_2\otimes L_{i_o}, \hs L_{i_o}=\begin{bmatrix}\ell_1/2 & 0 & \cdots & 0 \\ \ell_1 & \ell_2/2 & \ddots & \vdots \\ \vdots & \ddots & \ddots & 0 \\ \ell_1 & \cdots & \ell_{I-1} & \ell_I/2 \end{bmatrix},
	\]
	where we note that
	\[
	LC_\zeta=C_\zeta L, \hs LS_\zeta=S_\zeta L
	\]
	because $C_\zeta$ and $S_\zeta$ both are block diagonal matrix with the block diagonal element in the form of a constant times an identity matrix, $\ell_i$ is the length of the strut, the subscript $I$ is the half of the number of struts in the $i^{th}$ tensegrity radial, and the subscript $N$ is the total number of the radials in two wings, the last row in $L$ is for the mass center of body trunk, $x_0$, $y_0$ and $z_0$ are constant vectors composed of the coordinates of location where the tensegrity radials are connected to the body trunk and the coordinate of the mass center of body trunk (which is in the last row of $x_0$, $y_0$ and $z_0$), both expressed in body frame,
	\[
	C_\zeta=\mathrm{diag}(\cos\zeta_1, \cdots, \cos\zeta_{n-1}), \hs S_\zeta=\mathrm{diag}(\sin\zeta_1, \cdots, \sin\zeta_{n-1}),
	\]
	\[
	\Omega_\theta = C_\theta\otimes E_c + S_\theta \otimes E_s + I_{n-1}\otimes E_0
	\]
	\[
	C_\theta = \mathrm{diag}(\cos\theta_1, \cdots, \cos\theta_{n-1}), \hs S_\theta = \mathrm{diag}(\sin\theta_1, \cdots, \sin\theta_{n-1})
	\]
	\[
	E_c =\begin{bmatrix} 1 & 0 & 0 \\ 0 & 0 & 0 \\ 0 & 0 & 1 \end{bmatrix}, \hs E_s =  \begin{bmatrix} 0 & 0 & -1 \\ 0 & 0 & 0 \\ 1 & 0 & 0 \end{bmatrix}, \hs E_o = \begin{bmatrix} 0 & 0 & 0 \\ 0 & 1 & 0 \\ 0 & 0 & 0 \end{bmatrix}, \hs \Omega_\zeta=\begin{bmatrix} \Omega_{\zeta_1} \\ \vdots \\ \Omega_{\zeta_{n-1}} \end{bmatrix},
	\]
	$I_{n-1}$ is an identity matrix of dimension $n-1$, and
	\[
	J_2=\mathrm{diag}\left(\frac{m_1\ell_1^2}{12}, \cdots, \frac{m_{n-1}\ell_{n-1}^2}{12}\right), \hs J_3=\mathrm{diag}\left(\frac{m_1\ell_1^2}{12}, \cdots, \frac{m_{n-1}\ell_{n-1}^2}{12}\right),
	\]
	\[
	J_o = J_2\otimes I_{22} + J_3\otimes I_{33}, \hs
	I_{22}=\begin{bmatrix} 0 & 0 & 0 \\ 0 & 1 & 0 \\ 0 & 0 & 0 \end{bmatrix}, \hs I_{33}=\begin{bmatrix} 0 & 0 & 0 \\ 0 & 0 & 0 \\ 0 & 0 & 1 \end{bmatrix},
	\]
	where for a slender strut, the momentum of inertial $J_i$ in (\ref{eqn:J_vartheta}) is given by $J_i=\mathrm{diag}(0,J_{i_2},J_{i_3})$, $J_b$ is the moment of inertial around the mass center of body trunk alone axes of body frame.
	
	The simplifications in (\ref{eqn:Jtheta}), (\ref{eqn:eta_theta}) and (\ref{eqn:sigma_theta}) can be done as follows:
	\begin{align*}
		&\sum_{i=1}^n m_iQ(p_i^b - w^b)Q(p_i^b-w^b)^\intercal \cr
		=& \sum_{i=1}^n m_i \left(Q(p_i^b)-Q(w^b)\right)\left(Q(p_i^b)-Q(w^b)\right)^\intercal \cr
		=& \sum_{i=1}^n m_i Q(p_i^b)Q(p_i^b)^\intercal -m_oQ(w^b)Q(w^b)^\intercal \cr
		=&  Q(p^b)(M\otimes I_3)Q(p^b),
	\end{align*}
	and
	\begin{align*}
		&\sum_{i=1}^n m_i Q(p_i^b - w^b) \left(\frac{\partial p_i^b}{\partial\theta} - \frac{\partial w^b}{\partial\theta}\right)^\intercal \cr
		=& \sum_{i=1}^n m_i Q(p_i^b)\left(\frac{\partial p_i^b}{\partial\theta}\right)^\intercal  - m_oQ(w^b)\left(\frac{\partial w^b}{\partial\theta}\right)^\intercal \cr
		=& (y^\intercal MLC_\theta + z^\intercal MLS_\zeta S_\theta)\otimes e_1 -(x^\intercal MLC_\theta + z^\intercal MLC_\zeta S_\theta)\otimes e_2 \cr &
		+ (y^\intercal MLC_\zeta S_\theta - x^\intercal MLS_\zeta S_\theta)\otimes e_3,
	\end{align*}
	where we note that 
	\[
	\frac{\partial p^b}{\partial\theta} = \frac{\partial x}{\partial\theta} \otimes e_1^\intercal +\frac{\partial y}{\partial\theta} \otimes e_2^\intercal + \frac{\partial z}{\partial\theta} \otimes e_3^\intercal = -S_\theta C_\zeta\mathcal{L}^\intercal\otimes e_1^\intercal -S_\theta S_\zeta\mathcal{L}^\intercal\otimes e_2^\intercal + C_\theta\mathcal{L}^\intercal\otimes e_3^\intercal,
	\]
	\[
	w^b=\frac{1}{m_o}m\t p^b, \hs \frac{\partial w^b}{\partial\theta} = \frac{1}{m_o}\frac{\partial p^b}{\partial\theta}m,
	\]
	and
	\begin{align*}
		&\sum_{i=1}^n m_i\left(\frac{\partial p_i^b}{\partial\theta} - \frac{\partial w^b}{\partial\theta}\right)\left(\frac{\partial p_i^b}{\partial\theta} - \frac{\partial w^b}{\partial\theta}\right)^\intercal \cr
		=& \sum_{i=1}^n m_i \left(\frac{\partial p_i^b}{\partial\theta} \right)\left(\frac{\partial p_i^b}{\partial\theta} \right)^\intercal  - m_o\left(\frac{\partial w^b}{\partial\theta}\right)\left(\frac{\partial w^b}{\partial\theta}\right)^\intercal \cr
		=& S_\theta P_\zeta S_\theta + C_\theta Q C_\theta.
	\end{align*}
	Using the expression for $Q(p^b)$ given in (\ref{eqn:Qpb}), The expression of $J(\theta)$ in (\ref{eqn:Jtheta}) is further simplified as
	\begin{multline}\label{Jtheta_simplified}
		J(\theta)= z^\intercal M z\otimes(E_3E_3^\intercal) + y^\intercal My(E_2E_3^\intercal+E_3E_2^\intercal) + x^\intercal Mz \otimes(E_1E_3^\intercal + E_3E_1^\intercal) \cr + y^\intercal My\otimes E_2E_2^\intercal + x^\intercal My(E_1E_2^\intercal + E_2 E_1^\intercal) + x^\intercal Mx\otimes E_1E_1^\intercal + \Omega_\zeta^\intercal \Omega_\theta^\intercal J_o\Omega_\theta\Omega_\zeta + J_b 
	\end{multline}
	
	Next is to evaluate the matrix $G(\vartheta,\dot\vartheta)$. Let $\mathbb{J}(\vartheta)\dot\vartheta = \begin{bmatrix} \mu \\ \rho \end{bmatrix}$, the matrix $G(\vartheta,\dot\vartheta)$ is given by
	\[
	G(\vartheta,\dot\vartheta)  = \left(\frac{\partial\mathbb{J}(\vartheta)\dot\vartheta}{\partial\vartheta}\right)^\intercal - \frac{1}{2}\left(\frac{\partial\mathbb{J}(\vartheta)\dot\vartheta}{\partial\vartheta}\right) = \begin{bmatrix}\displaystyle\frac{\partial\mu}{\partial\phi} & \displaystyle \frac{\partial\rho}{\partial\phi} \\ \displaystyle\frac{\partial\mu}{\partial\theta} & \displaystyle \frac{\partial\rho}{\partial\theta} \end{bmatrix}^\intercal - \frac{1}{2}\begin{bmatrix} \displaystyle\frac{\partial\mu}{\partial\phi} &\displaystyle \frac{\partial\rho}{\partial\phi} \\\displaystyle \frac{\partial\mu}{\partial\theta} & \displaystyle\frac{\partial\rho}{\partial\theta} \end{bmatrix},
	\]
	where
	\[
	\mu = P(\phi)^\intercal J(\theta)P(\phi)\dot\phi + P(\phi)^\intercal\eta(\theta)\dot\theta, \hs \rho = \eta(\theta)^\intercal P(\phi)\dot\phi + \sigma(\theta)\dot\theta.
	\]
	
	The second term in (\ref{eqn:eom_theta}) is given as
	\[
	G(\vartheta,\dot\vartheta)\dot\vartheta = \begin{bmatrix} \displaystyle\left(\frac{\partial\mu}{\partial\phi}\right)^\intercal\dot\phi + \displaystyle \left(\frac{\partial\mu}{\partial\theta}\right)^\intercal\dot\theta \\ \displaystyle\left(\frac{\partial\rho}{\partial\phi}\right)^\intercal\dot\phi + \displaystyle \left(\frac{\partial\rho}{\partial\theta}\right)^\intercal\dot\theta \end{bmatrix} - \frac{1}{2}\begin{bmatrix} \displaystyle\frac{\partial\mu}{\partial\phi}\dot\phi+\displaystyle \frac{\partial\rho}{\partial\phi}\dot\theta \\\displaystyle \frac{\partial\mu}{\partial\theta}\dot\phi + \displaystyle\frac{\partial\rho}{\partial\theta}\dot\theta \end{bmatrix},
	\]
	The first block of $G(\vartheta,\dot\vartheta)\dot\vartheta$ is
	\begin{multline}
		\left(\frac{\partial\mu}{\partial\phi}\right)^\intercal\dot\phi + \left(\frac{\partial \mu}{\partial\theta}\right)^\intercal\dot\theta - \frac{1}{2}\left(\frac{\partial \mu}{\partial \phi}\dot\phi +  \displaystyle\frac{\partial\rho}{\partial\phi} \dot\theta\right) = \left(\frac{\partial\mu}{\partial\theta}\right)^\intercal\dot\theta + (A-A^\intercal)\dot\phi  \cr 
		+ P(\phi)^\intercal J_\theta \begin{bmatrix} \displaystyle\frac{\partial P(\phi)}{\partial\alpha}\dot\phi & \displaystyle\frac{\partial P(\phi)}{\partial\beta}\dot\phi & \displaystyle\frac{\partial P(\phi)}{\partial\gamma}\dot\phi \end{bmatrix}\dot\phi
	\end{multline}
	where
	\[
	A:= \begin{bmatrix} \displaystyle\frac{\partial P(\phi)^\intercal}{\partial\alpha}(J_\theta P(\phi)\dot\phi + \eta_\theta\dot\theta) & \displaystyle\frac{\partial P(\phi)^\intercal}{\partial\beta}(J_\theta P(\phi)\dot\phi + \eta_\theta\dot\theta) & \displaystyle\frac{\partial P(\phi)^\intercal}{\partial\gamma}(J_\theta P(\phi)\dot\phi + \eta_\theta\dot\theta) \end{bmatrix}
	\]
	
	
	\begin{multline}
		\left(\frac{\partial \mu}{\partial\theta}\right)^\intercal\dot\theta = 	P(\phi)^\intercal\left(2z^\intercal MLC_\theta\dot\theta\begin{bmatrix}\omega_1 \\ \omega_2 \\ 0\end{bmatrix} -2y^\intercal MLS_\zeta S_\theta\dot\theta\begin{bmatrix} \omega_1 \\ 0 \\ \omega_3 \end{bmatrix} - 2x^\intercal MLC_\zeta S_\theta\dot\theta\begin{bmatrix} 0 \\ \omega_2 \\ \omega_3 \end{bmatrix} \right. \\ +(-z^\intercal MLS_\zeta S_\theta\dot\theta + y^\intercal MLC_\theta\dot\theta)\begin{bmatrix} 0 \\ -\omega_3 \\ -\omega_2 \end{bmatrix} + (x^\intercal MLC_\theta\dot\theta - z^\intercal MLC_\zeta S_\theta\dot\theta)\begin{bmatrix} -\omega_3 \\ 0 \\ -\omega_1 \end{bmatrix} \\ \left. -(y^\intercal MLC_\zeta S_\theta\dot\theta + x^\intercal MLS_\zeta S_\theta\dot\theta)\begin{bmatrix} -\omega_2 \\ -\omega_1 \\ 0 \end{bmatrix}\right) \\ + 2C_\theta S_\theta  \left(b.d.\left(\begin{bmatrix} 1 & 0 & 0 \\ 0 & 0 & 0 \\ 0 & 0 & -1 \end{bmatrix}r.s.(\Omega_\zeta\omega) J_3\right)\right)^\intercal\dot\theta + (C_\theta^2 - S_\theta^2) \left(b.d.\left(\begin{bmatrix} 0 & 0 & 1 \\ 0 & 0 & 0 \\ 1 & 0 & 0 \end{bmatrix}r.s.(\Omega_\zeta\omega) J_3\right)\right)^\intercal\dot\theta \\ + P(\phi)^\intercal\begin{bmatrix}\cancelto{0}{\dot\theta^\intercal(S_\theta S_\zeta QC_\theta - C_\theta QS_\zeta S_\theta)\dot\theta} + (z^\intercal MLS_\zeta C_\theta - y^\intercal MLS_\theta)\dot\theta^2 \\ \cancelto{0}{\dot\theta^\intercal(C_\theta QC_\zeta S_\theta - S_\theta C_\zeta QC_\theta)\dot\theta} + (x^\intercal MLS_\theta - z^\intercal MLC_\zeta C_\theta)\dot\theta^2 \\ \cancelto{0}{\dot\theta^\intercal S_\theta(S_\zeta QC_\zeta - C_\zeta QS_\zeta)S_\theta\dot\theta} + (y^\intercal MLC_\zeta -x^\intercal MLS_\zeta)C_\theta\dot\theta^2 \end{bmatrix}
	\end{multline}
	where $S_\theta S_\zeta QC_\theta - C_\theta QS_\zeta S_\theta$, $C_\theta QC_\zeta S_\theta - S_\theta C_\zeta QC_\theta$ and $S_\zeta QC_\zeta - C_\zeta QS_\zeta$ are skew-symmetric matrices and $\dot\theta^\intercal S\dot\theta=0$ if $S$ is skew-symmetric, the operation $b.d.(a)$ is to make a block diagonal matrix with vector $a$ on the diagonal and the operation $r.s.(a)$ is to reshape the vector $a$ to a 3 by $n-1$ matrix.
	
	The second block of $G(\vartheta,\dot\vartheta)\dot\vartheta$ is
	\begin{multline}
		\left(\frac{\partial\rho}{\partial\phi}\right)^\intercal\dot\phi + \left(\frac{\partial\rho}{\partial\theta}\right)^\intercal \dot\theta - \frac{1}{2}\left(\frac{\partial \mu}{\partial\theta}\dot\phi + \frac{\partial\rho}{\partial\theta}\dot\theta\right) = \eta(\theta)^\intercal \begin{mat}{ccc}\displaystyle\frac{\partial P(\phi)}{\partial\alpha}\dot\phi & \displaystyle\frac{\partial P(\phi)}{\partial\beta}\dot\phi & \displaystyle\frac{\partial P(\phi)}{\partial\gamma}\dot\phi \end{mat}\dot\phi \\ + 2(S_\theta S_\zeta QC_\theta - C_\theta Q S_\zeta S_\theta)\dot\theta\omega_1 + 2(C_\theta QC_\zeta S_\theta - S_\theta C_\zeta QC_\theta)\dot\theta\omega_2 + 2S_\theta(S_\zeta QC_\zeta -C_\zeta QS_\zeta)S_\theta\dot\theta\omega_3 \\ + (C_\theta PS_\theta- S_\theta QC_\theta)\dot\theta^2 - \frac{1}{2}\frac{\partial P(\phi)J_\theta P(\phi)\dot\phi}{\partial\theta}\dot\phi
	\end{multline}
	where
	\begin{multline}
		\frac{\partial P(\phi)^\intercal J(\theta) P(\phi)\dot\phi}{\partial\theta} \dot\phi =
		2C_\theta L^\intercal Mz(\omega_1^2+\omega_2^2) -2 (C_\theta L^\intercal My - S_\theta S_\zeta L^\intercal Mz)(\omega_2\omega_3) \cr - 2(C_\theta L^\intercal Mx-S_\theta C_\zeta L^\intercal Mz)(\omega_1\omega_3) -2S_\theta S_\zeta L^\intercal My(\omega_1^2+\omega_3^2) \cr + 2S_\theta(S_\zeta L^\intercal Mx + C_\zeta L^\intercal My)(\omega_1\omega_2) - 2S_\theta C_\zeta L^\intercal Mx(\omega_2^2+\omega_3^2) 
	\end{multline}
	
	\section{Evaluation of $\partial l(\theta)/\partial\theta$}\label{appendix:generalized_force}
	Similarly to the mass center of the struts and body trunk, the coordinates of strut ends and the mass center of body trunk in body frame, $g^b(\theta)$, are given by
	\begin{equation}\label{eqn:gb}
		g^b(\theta)=x_g\otimes e_1 + y_g\otimes e_2 + z_g\otimes e_3, 
	\end{equation}
	where
	\begin{equation}
		x_g=\mathcal{L}C_\zeta \cos\theta + x_0, \hs y_g= \mathcal{L}S_\zeta \cos\theta +y_0, \hs z_g=\mathcal{L}\sin\theta + z0,
	\end{equation}
	with
	\[
	\mathcal{L} = \begin{bmatrix}\mathcal{L}_1 &  &  \\  & \ddots &  \\  &  & \mathcal{L}_N \\ 0 & \cdots & 0 \end{bmatrix}, \hs \mathcal{L}_i = I_2\otimes \mathcal{L}_{i_o}, \hs \mathcal{L}_{i_o}=\begin{bmatrix}\ell_1 & 0 & \cdots & 0 \\ \ell_1 & \ell_2 & \ddots & \vdots \\ \vdots & \ddots & \ddots & 0 \\ \ell_1 & \cdots & \ell_{I-1} & \ell_I \end{bmatrix},
	\]
	where the last row of zeros in $\mathcal{L}$ corresponds to the mass center of body trunk whose coordinates are given by the last element in vectors $x_0$, $y_0$ and $z_0$.

	The $i^{th}$ cable length $l_i(\theta)$ is written as
	\[
	l_i(\theta) = \sqrt{s_{i_x}^2 + s_{i_y}^2 + s_{i_z}^2}
	\]
	where $s_{i_x}$, $s_{i_y}$ and $s_{i_z}$ are components of cable vector $s_i(\theta)$ given by (\ref{eqn:cable_vector}). Then, the partial derivative of the cable length vector with respect to the body shape variable $\theta$ is given by
	\begin{align*}
		\frac{\partial l_i}{\partial\theta} &= \frac{\partial s_{i_x}}{\partial\theta}\frac{\partial l_i}{\partial s_{i_x}} + \frac{\partial s_{i_y}}{\partial\theta}\frac{\partial l_i}{\partial s_{i_y}} + \frac{\partial s_{i_z}}{\partial\theta}\frac{\partial l_i}{\partial s_{i_z}}\cr
		&=\frac{\partial s_{i_x}}{\partial\theta}\frac{s_{i_x}}{l_i} + \frac{\partial s_{i_y}}{\partial\theta}\frac{s_{i_y}}{l_i} + \frac{\partial s_{i_z}}{\partial\theta}\frac{s_{i_z}}{l_i} \hs \Rightarrow
	\end{align*}
	\begin{equation}\label{eqn:plpvartheta}
		\frac{\partial l}{\partial\theta} =\left(\frac{\partial s_x}{\partial\theta}\mathrm{diag}(s_x)+ \frac{\partial s_y}{\partial\theta}\mathrm{diag}(s_y) + \frac{\partial s_z}{\partial\theta}\mathrm{diag}(s_z)\right)\left(\mathrm{diag}(l)\right)^{-1}
	\end{equation}
	where $l$ is a vector composed of $l_i$'s, $s_x$, $s_y$ and $s_z$ are vectors composed of $s_{i_x}$, $s_{i_y}$ and $s_{i_z}$ ($i=1,\cdots,n_c$) which can be obtained from (\ref{eqn:cable_vector}) and (\ref{eqn:gb}):
	\[
	s_x = C_s x_g - x_0, \hs s_y = C_s y_g - y_0, \hs s_z = C_s z_y - z_0.
	\]
	Then, (\ref{eqn:plpvartheta}) becomes
	\begin{align}
		&\frac{\partial l}{\partial\theta} = \left(-S_\theta C_\zeta \mathcal{L}^\intercal C_s^\intercal \mathrm{diag}(s_x) - S_\theta S_\zeta \mathcal{L}^\intercal C_s^\intercal \mathrm{diag}(s_y) + C_\theta\mathcal{L}^\intercal C_s^\intercal \mathrm{diag}(s_z)\right)\left(\mathrm{diag}(l)\right)^{-1}
	\end{align}

	\section{Analytical fluid force model}\label{appendix:fluid_force}
\begin{figure}[!t]
		\centering
		\includegraphics[width=1.5in]{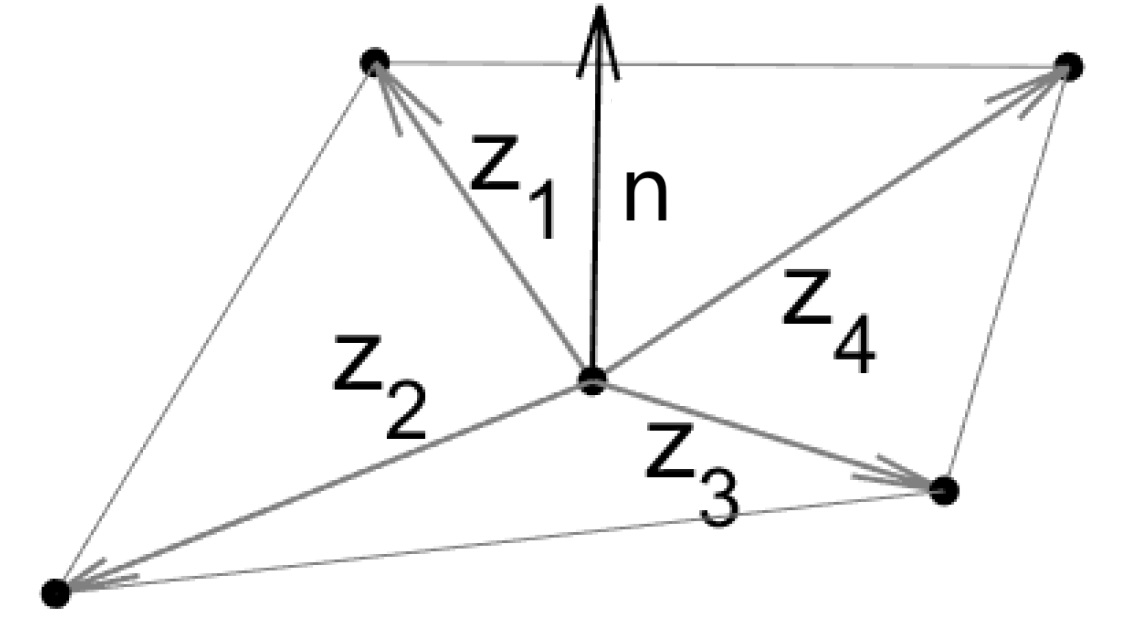}
		\caption{Calculation of the surface normal of the virtual fin surface at a node, adapted from \cite{chen:21}. The figure shows a node and its four surrounding nodes on the virtual fin surface. The normal $\mathsf{n}$ is approximated by the average of four triangular plane normals $\mathsf{n}=z/|z|$ where $z=(\sum_{i=1}^3z_i\times z_{i+1} + z_4\times z_1)/4$.}\label{fig:node_normal}
	\end{figure}

	To calculate the fluid force, the positions of strut ends supporting the top and bottom wing surfaces are averaged to form a single layer of nodes (we call it virtual wing surface to be referred to later). The velocities of this layer of nodes are used for fluid force calculation. The node velocity is obtained by taking the time derivative of (\ref{eqn:gi}) and then coordinatized in the body frame:
	\[
	\dot g_i^b(b) = \Omega(\phi)\dot w + Q(\omega)(g_i^b(\theta)-w^b(\theta)) + \left(\frac{\partial g_i^b}{\partial\theta}-\frac{\partial w^b}{\partial\theta}\right)^\intercal\dot\theta,
	\]
	and the velocities in the normal and tangential directions of wing local surface around individual nodes (Fig.~\ref{fig:node_normal}) are given by
	\[
	v_{n_i} = \mathsf{n}\mathsf{n}^\intercal \bar{\dot g}_i^b, \hs v_{t_i} = \bar{\dot g}_i^b - v_{n_i}
	\]
	where $\mathsf{n}$ is the normal of virtual wing surface expressed in the body frame (see Fig.~\ref{fig:node_normal}), and the overhead bar means the average velocity of strut ends on the top and bottom surfaces of the wing. For the body trunk, $\mathsf{n}=\begin{bmatrix} 0& 0 & 1\end{bmatrix}^\intercal$, and $\bar{\dot g}_i^b$ is the velocity of the mass center of the body trunk. The analytical fluid force is then given by \cite{liu:16,chen:21}:
	\begin{align} \nonumber
		f_{t_i}^b &=-0.5c_t\rho v_{t_i}|v_{t_i}|A_i, \\
		f_{n_i}^b &=-0.5c_n\rho v_{n_i}|v_{n_i}|A_i  \label{eqn:fluid}
	\end{align}
	where $f_{n_i}^b$ and $f_{t_i}^b$ are the fluid force vectors in the normal and tangential directions of wing local surface or the body trunk, $A_i$ is the surface area which, for wing local surfaces, equals half of the sum of four triangular plane areas (Fig.~\ref{fig:node_normal}) because each area is used twice by its vertices, $\rho$ is the fluid density, and the value for drag coefficient $c_t=0.097$ is obtained from \cite{fish:16} ($c_t$ during the steady-state swimming equals the ratio of thrust over $0.5\rho U^2S$, where $U$ is the swim speed and $S$ is the batoid planform area) and $c_n=1.18$ is adopted from \cite{chen:21}. The inertial component of fluid force is ignored in this study. The fluid force vector $f_i^b$ in (\ref{eqn:generalized_fluid_force}) is then given by
	\begin{equation}
		f_i^b=f_{n_i}^b+f_{t_i}^b
	\end{equation}

	The fluid torque on the body trunk is also considered and calculated by integrating the infinitesimal fluid torque generated by the normal component of fluid force on infinitesimal segments of body trunk \cite{jchen:nonlinear_dynamics}:  
	\[
	\mathsf{t}_{trunk,k}^b = \frac{1}{12}\rho c_n \omega_{trunk,k} v_{c,k} l_{trunk}^3w_{trunk}
	\]
	where $k$ ($k=x,y,z$) are components along the axes of body frame, $v_c$ and $\omega_{trunk}$ are respectively the velocity of the mass center and the angular velocity of the body trunk, $l_{trunk}$ and $w_{trunk}$ are respectively the body trunk length (width, length) and width (length, thickness) for $x$ ($y$, $z$)-direction torque.
	
	\section{Model parameters}\label{sec:model_parameters}
	The coordinates for radial wing grids and wing thickness function are respectively from Table~4.2 and Table~4.3 of \cite{blair:thesis} for the cownose ray. The dimension is amplified so that the tip-to-tip span width is 1.1~m. The body length is 0.6~m and the thickest part of the wing is 0.096~m. The wing radial grids provided in \cite{blair:thesis} are interpolated to have seven tensegrity radials each wing and five tensegrity units for the longest radial as shown in Fig.~\ref{fig:config_whole_body}c. The model is assumed to have the same density as the water and neutrally buoyant. The mass of wing segment is equally allotted to two struts in each tensegrity unit. The stiffness of horizontal cables within each tensegrity unit is assumed to be proportional to the cross-sectional area of wing segment where the tensegrity unit locates while the stiffness of vertical cables is 20 times that of the horizontal ones. The stiffness at the thickest place of the wing is set to be 54~N/m so that the first mode natural frequency of the wing is around 1~Hz. All springs in the vertical cables are pre-stretched from their rest-length by 5\%, and the pre-stretch lengths of horizontal cables are calculated from equilibrium of individual tensegrity radials. The resistive drag coefficient in analytical fluid force model (\ref{eqn:fluid}) for both the body and wing surfaces is $c_t=0.097$ which is obtained from computational fluid dynamics simulation where the data-fitted wing kinematics is taken as the immersed moving boundaries in viscous fluid in \cite{fish:16}. The normal drag coefficient in (\ref{eqn:fluid}) is $c_n=1.18$ from \cite{chen:21}. The amplitude of actuation sinusoidal signals $u$ in (\ref{eqn:tau}) is set to be proportional to wing thickness where the cable is located, and it decreases from the thickest part of the wing toward the ray head and wing tip directions from 80\% of $l_{i_0}$ to 15\%. The power input by cable actuation is calculated by $\sum_i \left(\frac{1}{T_c}\int_0^{T_c} k_iu_i\dot{l}_i\;\mathrm{d}t\right)$ (refer to (\ref{eqn:tau})) , where $T_c$ is the cycle period, for calculation of propulsion efficiency.
	
	\section{Basic kinematics}
	We first derive the velocity of an arbitrary point on a rigid body and in this process we get the formula for the partial derive of a position vector with respect to the orientation angles ($\alpha$, $\beta$ and $\gamma$) of body frame. It will be used later in the derivation of dynamic model of the system. The position of an arbitrary point on a rigid body in the inertial frame is given by 
	\[
	r+\Omega(\psi)^\intercal s
	\]
	where $r$ is position vector of mass center expressed in the inertial frame and $s$ is the vector from the body mass center to an arbitrary point on the rigid body expressed in the body frame, $\Omega(\psi)$ is a rotation matrix (3$\times$3 orthogonal matrix), $\psi=\begin{bmatrix} \alpha & \beta & \gamma \end{bmatrix}^\intercal$ is the Euler angles that specify the orientation of the body frame with respect to the inertial frame. Below, we show that the velocity of an arbitrary point on the rigid body expressed in the inertial frame is given by
	\[
	\frac{d}{dt}\left(r+\Omega(\psi)^\intercal s\right) = \dot{r} + \Omega(\psi)^\intercal(\omega\times s),
	\]
	where $\omega=\begin{bmatrix} \omega_1 & \omega_2 & \omega_3 \end{bmatrix}^\intercal$ is the angular velocity of the rigid body, expressed in the body frame.
	
	Recall that the cross product
	\begin{equation}\label{eqn:Q_def}
		a\times b = Q(a)b=Q(b)^\intercal a, \hs Q(c) = \begin{bmatrix} 0 & -c_3 & c_2 \\ c_3 & 0 & -c_1 \\ -c_2 & c_1 & 0 \end{bmatrix}
	\end{equation}
	where $c_i$ ($i=1,2,3$) are components of vector $c$. Since $\Omega(\psi)$ is an orthogonal matrix,
	\[
	\Omega\Omega^\intercal = I \hs \Rightarrow \hs  \frac{\partial\Omega}{\partial\alpha}\Omega^\intercal + \Omega\frac{\partial\Omega^\intercal}{\partial\alpha} = 0 \hs \Rightarrow \hs \Omega\frac{\partial\Omega^\intercal}{\partial\alpha} = \text{skew symmetric}.
	\]
	Hence,
	\[
	\Omega\frac{\partial\Omega^\intercal}{\partial\alpha} = Q(p_\alpha) \hs \Rightarrow  \hs \frac{\partial\Omega^\intercal}{\partial\alpha} = \Omega^\intercal Q(p_\alpha)
	\]
	holds for some vector $p_\alpha\in\IR^3$ that depends on $\psi$. Similar derivations can be applied for the partial derivatives with respect to $\beta$ and $\gamma$. Thus we have
	\[
	\frac{\partial\Omega^\intercal}{\partial\alpha} = \Omega^\intercal Q(p_\alpha), \hs \frac{\partial\Omega^\intercal}{\partial\beta} = \Omega^\intercal Q(p_\beta), \hs \frac{\partial\Omega^\intercal}{\partial\gamma} = \Omega^\intercal Q(p_\gamma),
	\]
	for some $p_\alpha$, $p_\beta$, $p_\gamma\in\IR^3$ depending upon $\psi$. Using these relations,
	\begin{align}
		\left(\frac{\partial\Omega(\psi)^\intercal s}{\partial\psi}\right)^\intercal &= \begin{bmatrix} \displaystyle \frac{\partial\Omega^\intercal s}{\partial\alpha} & \displaystyle \frac{\partial\Omega^\intercal s}{\partial\beta} & \displaystyle \frac{\partial\Omega^\intercal s}{\partial\gamma} \end{bmatrix} \cr
		&=\begin{bmatrix} \Omega^\intercal Q(p_\alpha)s & \Omega^\intercal Q(p_\beta)s & \Omega^\intercal Q(p_\gamma)s \end{bmatrix} \cr
		&=\begin{bmatrix} \Omega^\intercal Q(s)^\intercal p_\alpha & \Omega^\intercal Q(s)^\intercal p_\beta & \Omega^\intercal Q(s)^\intercal p_\gamma \end{bmatrix} \cr
		& = \Omega(\psi)^\intercal Q(s)^\intercal P(\psi) \label{eqn:partial_psi}  
	\end{align}
	where
	\begin{equation}\label{eqn:P_phi}
		P(\psi) = \begin{bmatrix} p_\alpha(\psi) & p_\beta(\psi) & p_\gamma(\psi) \end{bmatrix}.
	\end{equation}
	The velocity of an arbitrary point in the inertial frame is given by
	\begin{align*}
		\frac{d}{dt}\left(r+\Omega(\psi)^\intercal s\right) &= \dot r +\left(\frac{\partial \Omega^\intercal s}{\partial\psi}\right)^\intercal \dot\psi \\
		&= \dot r + \Omega(\psi)^\intercal Q(s)^\intercal P(\psi)\dot\psi \\ 
		&=\dot{r} + \Omega(\psi)^\intercal Q(s)^\intercal\omega \\
		&= \dot r + \Omega(\psi)^\intercal (\omega\times s)
	\end{align*}
	where $\omega=P(\psi)\dot\psi$
	is the angular velocity of the rigid body expressed in the body frame.
	
	For 3-2-1 Euler angles $\psi=\begin{bmatrix} \alpha & \beta & \gamma \end{bmatrix}^\intercal$, $P(\psi)$ and $\Omega(\psi)$ are given by
	\begin{align*}
		\Omega(\psi) &=\begin{bmatrix} 1 & 0 & 0 \\ 0 & c_\gamma & s_\gamma \\ 0 & -s_\gamma & c_\gamma \end{bmatrix} \begin{bmatrix} c_\beta & 0 & -s_\beta \\ 0 & 1 & 0 \\ s_\beta & 0 & c_\beta \end{bmatrix} \begin{bmatrix}  c_\alpha & s_\alpha & 0 \\ -s_\alpha & c_\alpha & 0 \\ 0 & 0 & 1 \end{bmatrix} \\
		&=\begin{bmatrix} c_\alpha c_\beta & s_\alpha c_\beta & -s_\beta \\ -s_\alpha c_\gamma + c_\alpha s_\beta s_\gamma & c_\alpha c_\gamma + s_\alpha s_\beta s_\gamma & c_\beta s_\gamma \\ s_\alpha s_\gamma + c_\alpha s_\beta c_\gamma & -c_\alpha s_\gamma + s_\alpha s_\beta c_\gamma & c_\beta c_\gamma \end{bmatrix}
	\end{align*}
	\begin{equation}\label{eqn:P_phi_321}
		P(\psi)=\begin{bmatrix} -s_\beta & 0 & 1 \\ c_\beta s_\gamma & c_\gamma & 0 \\ c_\beta c_\gamma & -s_\gamma & 0 \end{bmatrix}
	\end{equation}
	where $s$ and $c$ are abbreviations of $\sin$ and $\cos$ functions.

\end{document}